\documentclass[12pt,preprint]{aastex}

\shorttitle{RS Oph: First Science with the Keck Interferometer Nuller}
\shortauthors{R. K. Barry et al.}

\begin{document}

\title{Milliarcsecond N-Band Observations of the Nova RS Ophiuchi: \\
     First Science with the Keck Interferometer Nuller}

\author{ R. K. Barry\altaffilmark{1,2}, W. C. Danchi\altaffilmark{1}, W. A. Traub\altaffilmark{3}, J. L. Sokoloski\altaffilmark{4}, J. P. Wisniewski\altaffilmark{1}, E. Serabyn\altaffilmark{3}, M. J. Kuchner\altaffilmark{1}, 
R. Akeson\altaffilmark{5}, E. Appleby\altaffilmark{6}, J. Bell\altaffilmark{6}, A. Booth\altaffilmark{3}, H. Brandenburg\altaffilmark{5}, M. Colavita\altaffilmark{3}, S. Crawford\altaffilmark{3}, M. Creech-Eakman\altaffilmark{3}, W. Dahl\altaffilmark{6}, C. Felizardo\altaffilmark{5}, J. Garcia\altaffilmark{3}, J. Gathright\altaffilmark{6}, M. A. Greenhouse\altaffilmark{1}, J.Herstein\altaffilmark{5}, E. Hovland\altaffilmark{3}, M. Hrynevych\altaffilmark{6}, C. Koresko\altaffilmark{3}, R. Ligon\altaffilmark{3}, B. Mennesson\altaffilmark{3}, R. Millan-Gabet\altaffilmark{5}, D. Morrison\altaffilmark{6}, D. Palmer\altaffilmark{3}, T. Panteleeva\altaffilmark{6}, S. Ragland\altaffilmark{6}, M. Shao\altaffilmark{3}, R. Smythe\altaffilmark{3}, K. Summers\altaffilmark{6}, M. Swain\altaffilmark{3}, K. Tsubota\altaffilmark{6}, C. Tyau\altaffilmark{6}, G. Vasisht\altaffilmark{3}, E. Wetherell\altaffilmark{6}, P. Wizinowich\altaffilmark{6}, J. Woillez\altaffilmark{6} }

\altaffiltext{1}{NASA Goddard Space Flight Center, Exoplanets and Stellar Astrophysics Laboratory,
Greenbelt, MD 20771}\email{Richard.K.Barry@nasa.gov}
\altaffiltext{2}{Department of Physics and Astronomy, The Johns Hopkins University, Baltimore, MD 21218}
\altaffiltext{3}{Jet Propulsion Laboratory, California Institute of Technology, Pasadena, CA 91109}
\altaffiltext{4}{Columbia University, Department of Physics, NY 10027 }
\altaffiltext{5}{Michelson Science Center, Caltech 100-22 Pasadena, CA 91125}
\altaffiltext{6}{W. M. Keck Observatory, California Association for Research in Astronomy, 65-1120 Mamalahoa Highway, Kamuela, HI 96743}
\begin{abstract}

We report observations of the nova RS Ophiuchi (RS Oph) using the Keck Interferometer Nuller 
(KIN), approximately 3.8 days following the most recent outburst that occurred on 2006 February
12.  These observations represent the first scientific results from the KIN, which operates in N-band from 8 to 12.5 $\mu$m in a nulling mode.    The nulling technique is the sparse aperture equivalent of the conventional coronagraphic technique used in filled aperture telescopes.  In this mode the stellar light itself is suppressed by a destructive fringe, effectively enhancing the contrast of the circumstellar material located near the star.  By fitting the unique KIN  data, we have obtained an angular size of the mid-infrared continuum of 6.2, 4.0, or 5.4 mas for a disk profile, gaussian profile (FWHM), and shell profile respectively.  The data show evidence of enhanced neutral atomic hydrogen emission and atomic metals including silicon located in the inner spatial regime near the white dwarf (WD) relative to the outer regime.  There are also nebular emission lines and evidence of hot silicate dust in the outer spatial region, centered at $\sim$ 17 AU from the WD, that are not found in the inner regime.  Our evidence suggests that
these features have been excited by the nova flash in the outer spatial regime before the blast wave reached these regions.  These identifications support a model in which the dust appears to be present between outbursts and is not created during the outburst event.  We further discuss the present results in terms of a unifying model of the system that includes an increase in density in the plane of the orbit of the two stars created by a spiral shock wave caused by the motion of the stars through the cool wind of the red giant star. These data show the power and potential of the nulling technique which has been developed for the detection of Earth-like planets around nearby stars for the Terrestrial Planet Finder Mission and Darwin missions.

\end{abstract}

\keywords{novae: cataclysmic variables  --- stars: dwarf novae  ---
stars: individual(\objectname{RS Ophiuchi},
\object{HD162214})--- techniques: interferometric--- techniques: high angular resolution }

\section{INTRODUCTION}

Classical novae (CN) are categorized as cataclysmic variable stars that have had only one \textit{observed} outburst - an occurrence typified by a Johnson V-band brightening of between six and nineteen magnitudes \citep{war95}.  These eruptions are well-modeled as thermonuclear runaways (TNR) of hydrogen-rich material on the surface of white dwarf (WD) primary stars that, importantly, remain intact after the event.  Current theory tells us that CN can be modeled as binary systems in which a lower-mass companion -- the secondary -- orbits the WD primary such that the rate of mass transfer giving rise to the observed eruption is very low.  Recurrent novae (RN) are a related class of CN which have been observed to have more than one eruption.  Like CN, RN events are well-represented as surface TNR on WD primary stars in a binary system, but are thought to have much higher mass transfer rates commensurate with their greater eruption frequency.  There are two types of systems that produce recurrent novae -- CVs, in which the WD accretes from a main sequence star that orbits the WD on a time scale of hours, and symbiotic stars, in which the WD accretes from a red-giant companion that orbits that WD on a time scale of years.

CN and RN produce a few specific elemental isotopes by the entrainment of metal-enriched surface layers of the WD primary during unbound TNR outer-shell fusion reactions.  In contrast, type Ia supernovae produce most of the elements heavier than helium in the Universe through fusion reactions leading to the complete destruction of their WD primary.  Some theoretical models indicate that RN could be a type of progenitor system for supernovae.  Importantly, these theories are predicated on two critical factors: 1.) the system primary must be a compact carbon-oxygen core supported solely by electron degeneracy pressure and 2.) there must be some mechanism to allow the WD mass to increase secularly towards the Chandrasekhar limit.  

The nova RS Oph has undergone six recorded episodic outbursts of irregular interval in 1898 \citep{fle04}, 1933 \citep{ada33}, 1958 \citep{wal58}, 1967 \citep{bar69}, 1985 \citep{mor85} and now 2006.  There are also two possible outbursts in 1907 \cite[]{sch04} and 1945 \cite[]{opp93}.  All outbursts have shown very similar light curves.  This system is a single-line binary, symbiotic with a red giant secondary characterized as K$5.7 \pm 0.4$  I-II \citep{ken87} to a K7 III \citep{mur99} in quiescence and a white dwarf primary in a $455.72\pm0.83$ day orbit about their common center of mass as measured using single-line radial velocity techniques \cite[]{fek00}.  \citeauthor{opp93} examined all the outbursts and found that V band luminosity of RS Oph decreased averaged 0.09 magnitudes/day for the first 43 days after outburst. A 2-magnitude drop would then require on average 22 days, establishing RS Oph as a ÒfastÓ nova based on the classification system of \cite{pay57}.  
 
The most recent outburst of the nova RS Oph was discovered at an estimated V-band magnitude of 4.5 by H. Narumi of Ehime, Japan on 2006 February 12.829 UT \cite[]{nar06}.  This is 0.4 mag brighter than its historical average AAVSO V-band {\it peak} magnitude so it is reasonable to take Feb 12.829 (JD 2453779.329) as day zero.  The speed of an outburst is characterized by its $t_2$ and $t_3$ times
which are  the intervals in days from the visible maximum until the system has dimmed by 2 and 3 magnitudes, respectively.   For this outburst the $t_2$ and $t_3$ times are 4.8 and 10.2 days, respectively.

The distance to the RS Oph system is of  importance to the interferometry community as it  effects interpretation of astrometric data (cf. \citet{mon06}).  There has been a good deal of disagreement   in the literature with a surprisingly broad range, from as near as  0.4 kpc \citep{hac01} to as far as 5.8 kpc \citep{pot67}.  
 \citet{bar08} have recently undertaken a thorough review of the various techniques that have been used to derive a distance to RS Oph  and obtain a distance  of $1.4^{+ 0.6}_{-0.2} $ kpc.
 It is this value that we adopt for  astrometric calculations in this paper. 

  

The structure of this paper is as follows. We report high-resolution N-band observations of  RS Oph using the nulling mode of the 85 meter baseline Keck Interferometer, beginning with a discussion of
the nulling mode itself in Section 2.  We discuss the observations in Section 3, and the data and
analysis in Section 4.   In Section 5 we  introduce a new physical model of the system, 
which unifies many of the observations into a coherent framework.  The results of this paper and those of 
other recent observations of RS Oph are discussed in the context of this model in Section 6.
Finally, Section 7 contains a summary of our major results and conclusions.  

 
\section{THE KECK INTERFEROMETER NULLER}

The KIN is designed to detect faint emission due, e.g., to an
optically-thin dust envelope, at small angular distances from a bright
central star (Serabyn, Colavita and Beichman 2000). Its operation
differs from a more common fringe scanning optical interferometer in
that a nulling stage precedes the fringe scanning stage (Serabyn et
al. 2004, 2005, 2006; Colavita et al. 2006). The basic measurement thus
remains the fringe amplitude, but both the meaning of the fringe
signal in relation to the source and the processing of the fringe
information differ from the normal case of a standard visibility
measurement.  Here we provide only a brief description of the
measurement process, because this has been and will be described in
depth elsewhere (Serabyn et al. 2004, 2005, 2006, Colavita et al. 2006;
Serabyn et al. 2008, in prep).

\begin{figure}
\begin{center}
\includegraphics[width=4.5in]{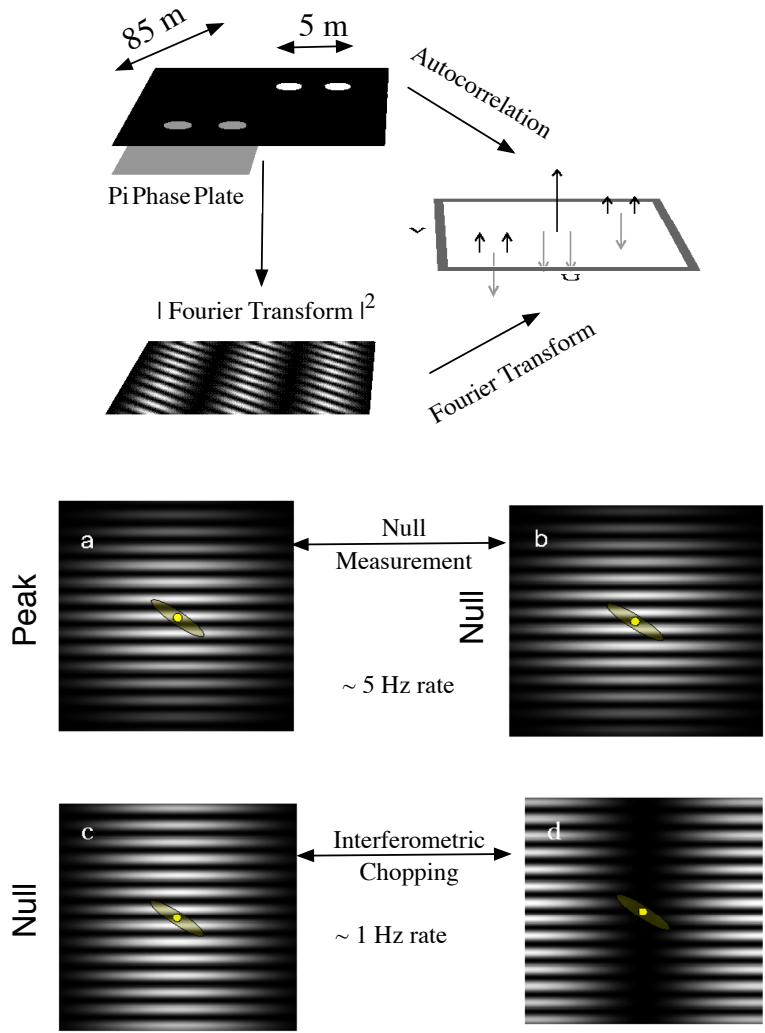}
\caption{The upper part of plot displays a conceptual view of  the operation of the KIN system.  There are two short baselines, separated by ~ 5 m, and two long baselines separated by ~ 85 m.  A $\pi$ phase shift is applied to the pupils on the long baseline.  The autocorrelation of the four pupils is shown as well as the equivalent intensity pattern, also called the transmission pattern.  Radiation passing through the white striped regions is detected by the system.  Lower part of the plot displays the sequence used for the measurement process.  There are two chopping sequences, (a) and (b) display the $\sim$ 5 Hz chopping between the null and bright fringe patterns on the source using the 85 m baseline for starlight subtraction, used to measure the null response, while (c) and (d) show the ~$\sim$1 Hz chop sequence between the two short baselines used to remove the telescope and sky backgrounds. Note that the long-baseline fringes are not to scale with the field of view in this depiction. \label{KINconcept}}
\end{center}
\end{figure}

To remove both the stellar signal and the thermal background in the
MIR, a two stage interferometer has been developed. 
To implement this approach, each Keck telescope is first
split into two half-apertures, to generate a total of four collecting
sub-apertures. The starlight is then first nulled on the two long (85
m) parallel baselines between corresponding Keck subapertures. This
generates the familiar sinusoidal fringe pattern on the sky (Fig. \ref{KINconcept}),
except that the central dark fringe on the star is achromatic, and
fixed on the star, to achieve deep and stable rejection of the
starlight. After the nulling stage, the residual, non-nulled light
making it through the first stage fringe pattern is measured by a
fringe scan in a second stage combiner, the ``cross-combiner'', which
combines the light across the Keck apertures ($\sim$ 4 m
baseline). Thus, what is measured is the fringe amplitude of this
``nulled source brightness distribution'' (Serabyn et al. 2008, in
prep.)  This quantity is then normalized by the total signal. This is
measured by moving the nullers to the constructive phase, and again
scanning the cross-combiners. The basic measured quantity, the null
depth, $N$, is then the ratio of the signals with the star in the
destructive and constructive states.  $N$ is related to the classical
interferometer visibility $V = (I_{max} -I_{min})/(I_{max} +I_{min})$,
the modulus of the complex visibility $\hat{V}$, by a simple formula:

\begin{equation}
N ={ {1-V}\over{1+V}} ~~ .
\end{equation}

The long nulling baselines produce fringes spaced at about 23.5 mas at
10 $\mu$m, while the short baseline produces fringes spaced at $ \sim
400$~mas,which is similar to the size of the primary beam and is
assumed to be large compared to the extent of the target object.
Modulating its phase therefore modulates the transmitted flux of the
entire astronomical source, as modified by the sinusoidal nuller
fringes, and so the amplitude of this modulated signal gives the flux 
that passes through the fringe transmission pattern
produced by the long baseline nullers.

\section{OBSERVATIONS}

We observed the nova about an hour angle of about -2.0 on the Keck Interferometer in nulling mode on 2006 Feb 16 with a total of three observations between day 3.831 and 3.846 post-outburst bracketed with observations of two calibrators stars, $\rho$ Boo and $\chi$ UMa.    Data were obtained at N-band 
(8 to 12.5 $\mu$m) through both ports of the KALI spectrograph with the gating of data for long baseline phase delay and group delay turned on.  The Infrared Astronomical Satellite (IRAS) Low Resolution Spectrometer (LRS) spectra for the calibrator stars were flux-scaled according to the broadband IRAS 12 $\mu$m fluxes.   The nova data were flux calibrated and telluric features were removed using calibrator
data.  These calibrators are well matched to the target flux and their size and point-symmetry are well known. A journal of observations is presented in Table \ref{observations}.

Our data analysis involves removing biases and coherently demodulating the short-baseline fringe with the long-baseline fringe tuned to alternate between constructive and destructive phases, combining the results of many measurements to improve the sensitivity, and estimating the part of the null leakage signal that is associated with the finite angular size of the central star. Comparison of the results of null measurements on science target and calibrator stars permits the instrumental leakage - the ``system null leakage'' - to be removed and the off-axis light to be measured. 


Sources of noise in the measurements made by this instrument have been well described elsewhere \citep{kor06}, however, we outline them here for reference.  The null leakage and intensity spectra include contributions from the astrophysical size of the object, phase and amplitude imbalances, wavefront error, beamtrain vibration, pupil polarization rotation, and pupil overlap mismatch. There are also biases, mostly eliminated by use of sky frames, in the calculated fringe quadratures, caused mainly by thermal background modulation due to residual movement of the mirrors used to shutter the combiner inputs for the long baselines. Another source of error is the KALI spectrometer channel bandpass, which is large enough to produce a significant mismatch at some wavelengths between the center wavelength and the short-baseline stroke OPD.  This effect, termed ÔwarpingÕ, distorts the quadratures and is corrected by a mathematical dewarping step accomplished during calibration. There is also the effect of the partial resolution of extended structure on the long baselines which will cause the flux to be undercounted by some amount depending on the spatial extent and distribution of the emitting region.  Compared to the error bars this is a rather small effect for most normal stars, and is unlikely to have much influence on the actual spectrum, though, unless the emission lines are coming from a very extended shell Ð approaching 25 mas in angular size.  We do not expect this to be the case for nova RS Oph at day 3.8.

The most important contributor to measurement noise is sky and instrument drifts between target and calibrator.
Other potential sources of noise include the difference between the band center wavelength for the interferometer and that of the IRAS LRS calibrator, undercounting of stellar flux resulting from glitches in the short-baseline phase tracking, and the chromaticity of the first maximum of the long-baseline fringe.  None of these are  significant for the following reasons. First, the calibration is based not on broadband photometry but on the IRAS LRS spectrum.  Second, short-baseline tracking glitches happen nearly as frequently on target observations as on calibrator observations so they should have minimal effect on overall calibrated flux. Third, the effect of chromaticity should  be negligible because fringe detection is done on a per-spectral-channel basis.  As a result, it is only affected by dispersion within the individual KALI spectral channels, which are about 0.3 $\mu$m in width.  

\section{DATA \& ANALYSIS}


We developed a mathematical solution and  software suite to model the observatory and source brightness distribution.  We used this suite to conduct an exhaustive grid search and to 
generate a monte carlo confidence interval analysis of solution spaces of  these models.  We explored three types of models for the source surface brightness distribution; Gaussian, disk, and shell.  Limited (u,v) coverage permitted only rotationally-symmetric models with two parameters - size and flux.  We used $\chi^2$ minimization to obtain the best fit models for both the inner and outer spatial regimes simultaneously.  Table \ref{models} displays size measurements, flux values, and  one-$\sigma$ confidence interval values. The error bars have been increased slightly beyond the $\chi^2$ values
to incorporate the effect of an adopted 0.005 one-sigma systematic error which would be correlated among measurements at different wavelengths.

The measurements made with the two KALI ports are somewhat independent - the data they produce are combined for purposes of fringe tracking, but not for data reduction.  The system null and the final calibrated leakage are computed separately for the two ports.  The apparent inconsistency detected between the ports is the result of optical alignment drifts at the time of the measurement.  In particular, the last calibrator measurement showed a sudden change in the system null for Port 1, while for all the other calibrator measurements the system nulls were stable.  We therefore compared our source brightness distribution models against KALI port 2 data alone.  For the best-fit models in Table \ref{models} we used {\it Spitzer} spectra to identify and remove emission features centered at 8.7, 9.4, 10.4, 11.4, and 12.5 $\micron$ in the KIN inner and 8.9, 9.8, and 11.4 $\micron$ in the KIN outer spatial regime.  We removed the emission feature data because our intention was to model the continuum. 

\begin{figure}
\begin{center}
\includegraphics[width=6.5in]{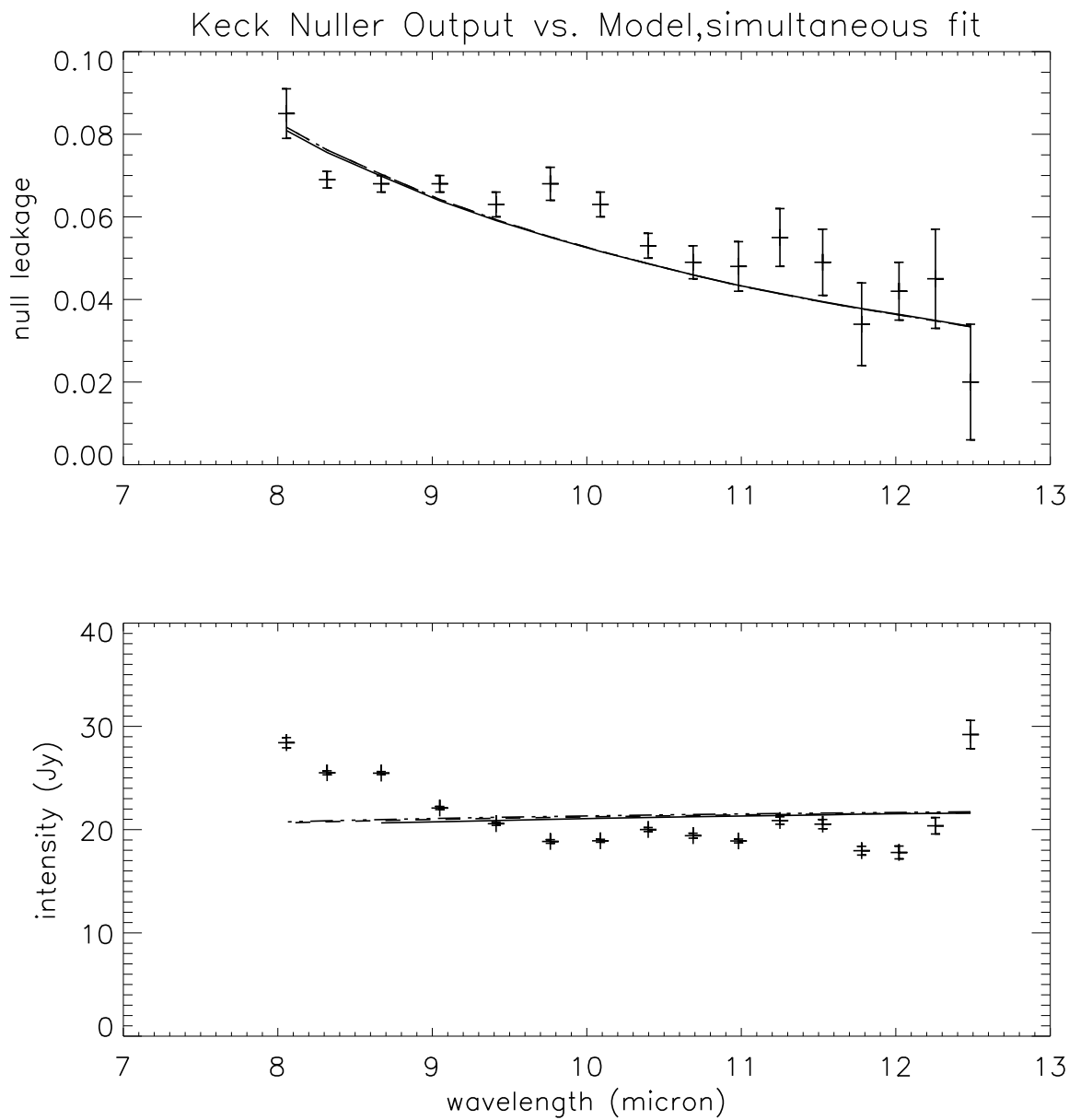}
\caption{Plot of three best-fit continuum models against {\it Keck} Nuller data.  All three models; disk, shell, and Gaussian, that minimized $\chi^{2}$ simultaneously against the null leakage and intensity spectra, lie effectively on top of one another. The top trace gives the dimensionless null leakage or interferometric observable which is the null fringe output divided by the intensity spectrum.  The lower trace is the constructive fringe output or intensity spectrum.  As described in the text we have removed several data points associated with emission features from the original set for the purpose of fitting the continuum.  \label{modelfit}}
\end{center}
\end{figure}

Figure \ref{modelfit} shows two sets of $8 - 12.5$ $\mu$m spectra of RS Oph on day 3.8 post-outburst.  The upper plot shows the outer spatial regime, which is the  dimensionless null leakage spectrum, i.e.,  the intensity of light remaining after destructive interference divided by the intensity spectrum,  plotted against wavelength in microns.  The lower plot is  the intensity spectrum, which is light principally from the inner 25 mas centered on the source brightness distribution orthogonal to the Keck Interferometer baseline direction - 38 degrees East of North.  The null leakage spectrum may be broadly described as a distribution that drops monotonically with increasing wavelength overlaid with wide, emission-like features.  The intensity spectrum, with an average flux density of about 22 Jy over the instrument spectral range, may be similarly described but with a continuum that has a saddle shape with a distinct rise at each end.   The underlying shape curves  upward for wavelengths shorter than approximately 9.7 $\mu$m and longer than 12.1 $\mu$m.    Overlaid on each of these are traces representing simple models of source brightness distributions fit to the data, described above.



\begin{figure}
\begin{center}
\includegraphics [width=5.5in]{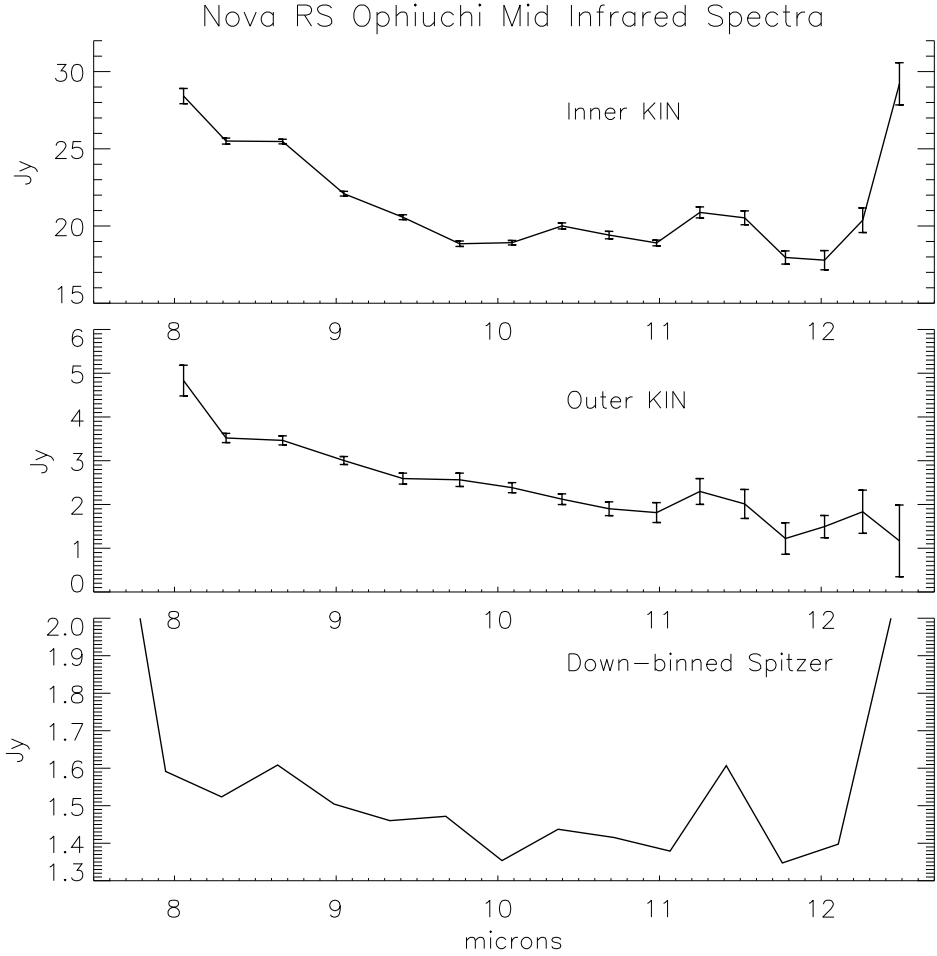}
\caption[KIN data compared to \textit{Spitzer} data]{Plot of KIN spectra together with a {\it Spitzer} spectrum. Upper trace is the intensity spectrum or inner KIN spatial regime - light dominated by the inner 25 mas about the center of the source brightness distribution at mid band.  The middle trace is the nulled fringe output which is the interferometric observable times the intensity spectrum multiplied by 2 (to correct for transmission through the fringe pattern for extended sources) to give the source brightness in the outer spatial regime.  This is predominantly emission from material greater than about 12.5 mas ($\sim$17 AU at 1.4 kpc) from the center of the source.  The lower trace is {\it Spitzer} data \citep{eva07b} from day 63 boxcar averaged to yield approximately the same spectral resolution as the {\it Keck} Interferometer Nuller.  None of the data were continuum normalized. Note the features between 9 and 11 $\micron$ and how the inner and outer spatial regime spectra are different from one another.  Note that the Spitzer spectra have very low spatial resolution and combine the light from the entire region detected by both KIN inner and outer spatial regimes.}
\label{kn_spectra} 
\end{center}
\end{figure}

Figure \ref{kn_spectra} shows the inner and outer spatial regimes from the KIN data together with {\it Spitzer} data from day 63 \citep{eva07b} .  The absolute fluxes of these data are unscaled and given with a broken ordinate axis for clarity.  
The outer KIN spectrum flux has been multiplied by 2 to correct
for the transmission through the fringe pattern for extended
sources.
In efforts to identify the sources of these emission features we took the high resolution {\it Spitzer} spectra and, using boxcar averaging, re-binned the data until it and the KIN data had equivalent resolution.  The binned {\it Spitzer}  spectrum is quite similar in character to the KIN spectra.  Importantly, the sum of the inner and outer spatial regime KIN spectra is  nearly identical to the binned {\it Spitzer} spectrum with the exception that the absolute scale magnitude is, on day 3.8, about an order of magnitude greater than that of the {\it Spitzer} spectrum.  This is as expected because
the measured flux drops with time after the peak due to cooling and the summed inner and outer spatial regime data should very nearly reproduce the transmission of a filled aperture telescope of equivalent diameter.  

The wavelength range sampled by the KALI spectrometer, 8 - 12.5 $\mu$m, covers many important discrete transitions including molecular rotation-vibration, atomic fine structure, and electronic transitions of atoms, molecules, and ions. This range also samples several important transitions in solids such as silicates found in dust and polycyclic aromatic hydrocarbons (PAHs).  With the exception noted below, spectral features in the KIN data are clearly not resolved by the instrument and are often, in the associated {\it Spitzer} spectrum, Doppler broadened and blended.  Also, because the {\it Spitzer} spectra were not taken contemporaneously with the KIN spectra and because of the  transient nature of the RN, identification of KIN features with
{\it Spitzer} lines is necessarily tentative.  

\begin{figure}
\begin{center}
\includegraphics[width=6.5in]{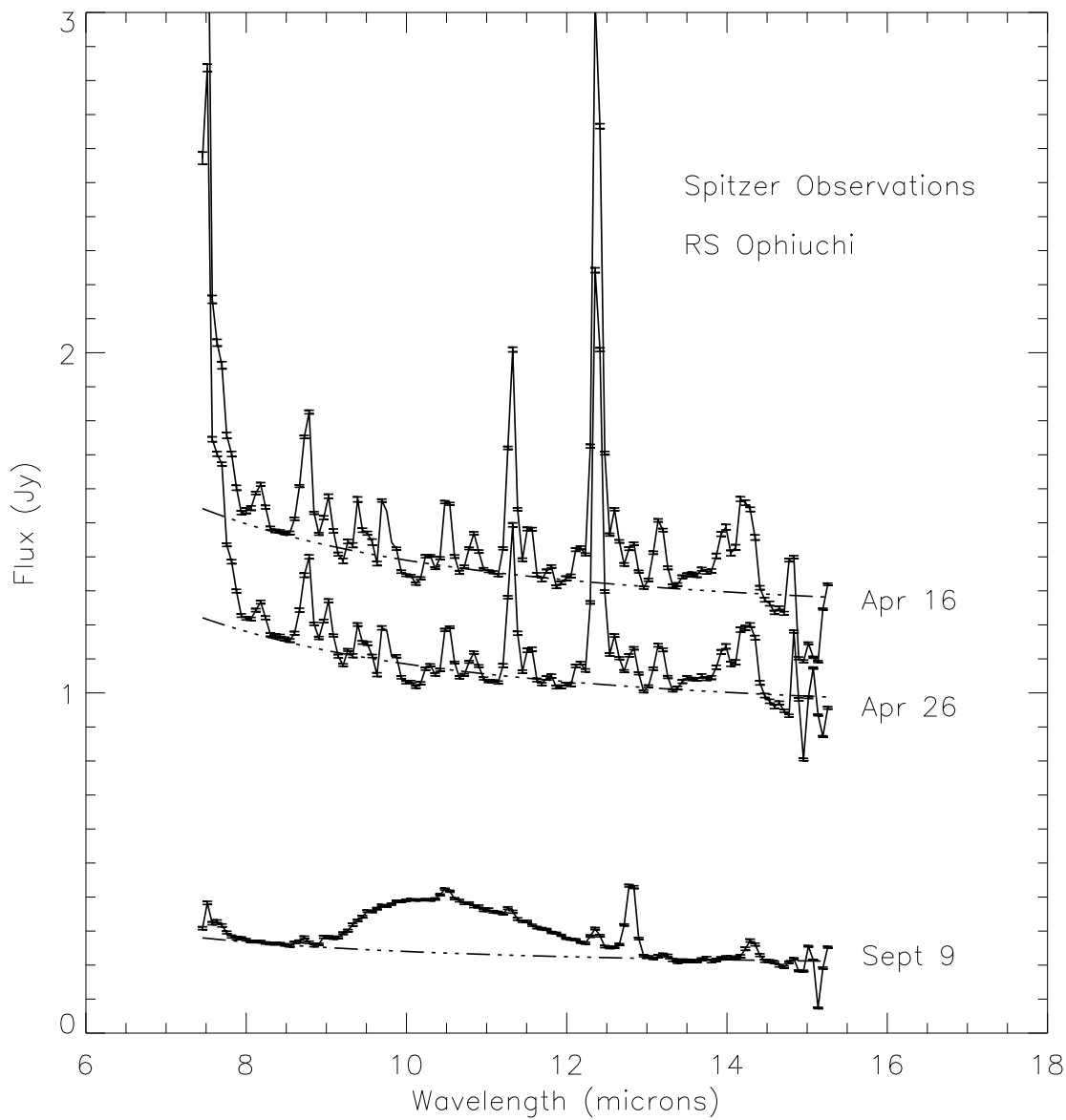}
\caption{{\it Spitzer} space telescope spectra from days 63, 73 and 209.  Here we see that the continuum drops rapidly as time advances with the spectral emission features almost identical on days 4/16 and 4/26.  Data obtained on 9/9 is starkly different with a strong solid-state feature evident between 9 and 13 microns (see \citet{eva07c} for more details on the {\it Spitzer} data). \label{continuum}}
\end{center}
\end{figure}

Figure \ref{continuum} shows {\it Spitzer} spectra  on days 63, 73 and 209 after peak V-band brightness.  The continuum decreases monotonically with time.  The  continuum values  are approximately 1.4, 1.1, and 0.25 Jy for  April 14, 26 and September 9, respectively.  The first two spectra on  April 14 and 16 show strong atomic lines with no obvious evidence of dust emission.  However it is expected that thermal bremsstrahlung from the central source will overwhelm any other faint sources. Additionally, there are several narrow emission features in this latter spectrum  which appear to be similar to those in the earlier spectra.   In contrast, the spectrum taken on September 9 shows a distinct broad emisison feature  from 8.9 to 14.3 $\mu$m peaking at 10.1 $\mu$m.  We proceed on the assumption that this is  emission from dust in the vicinity of the nova.  

Referring again to the KIN spectrum in Figure \ref{kn_spectra},  strong continuum radiation is apparent in both inner and outer spatial regimes.  In contrast, while the continuum was still clearly visible in J, H, I, K bands on February 24 and detectable on April 9 \citep{eva07a}, it has subsided by day 63 in Figure \ref{kn_spectra} and Figure \ref{continuum}.  Comparing the spectra from RS Oph to those from V1187 Sco \citep{lyn06} note that the continuum given for V1187 Sco is non-Planckian showing an excess longwards of 9 $\mu$m and is strongly red as compared to a F5V Kurucz spectrum.   The V1187 Sco and RS Oph continua have slopes that agree to within 10\%.  While both free-bound and free-free transition processes lead to emission of continuum radiation, in the MIR spectral range thermal bremsstrahlung free-free emission dominates.  We attribute the drop in continuum radiation to the transition to line-emission cooling mechanisms.  Also, by the time {\it Spitzer} data  were taken, the object had become less dense and so the emission coefficient for thermal bremsstrahlung (proportional to number density of protons and electrons) had dropped considerably.  The continuum emission is described in detail by \citet{bar08b}. 

Table \ref{spitzlinelist} gives identification of all narrow features in Figure \ref{continuum} where it is possible to do so, as follows.  We generated line lists of atomic species assuming that all ionization stages of all elements were possible for the three {\it Spitzer} spectra.  We assumed  standard cosmic abundances \citep{gre84}.  After continuum normalization, each emission line was fitted with a Gaussian and, where necessary, de-blended using IRAF.  Our fitted emission features were compared with 
the {\it Spitzer} list.

Table \ref{kinfeaturelist} gives our identification of particular emission sources with the continuum-normalized spectral features in the KIN inner and outer spectra.  The center wavelength of each of the broad features is listed in the first column, while the second column displays a width for each feature defined as the cutoff wavelength at the intersection of the the feature and the unity continuum line - the full-width zero intensity (FWZI) level of the feature.  The flux and the one-$\sigma$ uncertainty
contained in that feature through measurement of the total area under it and above the unity continuum level is in the third column.
The fourth column shows the particular atomic species found in the {\it Spitzer} spectrum with each KIN spectral feature.    Because  {\it Spitzer}  lacks the spatial resolution to discriminate the inner from the outer regions of the nova and the KIN lacks the spectral resolution to discriminate among atomic species we assumed that features in the KIN spectra would most reasonably be identified with {\it Spitzer} features that were in later spectra.  In particular,  {\it Spitzer} emission lines features identified with corresponding ones in KIN inner spatial regime, dominated by light originating in the close vicinity the WD, are well-represented by a cosmic distribution of atomic elements. We assumed that any condensates within the blast radius of the nova would be sublimated away and dissociated into atoms, and that any nucleosynthesis that occurred in the outer layers of the WD during TNR would negligibly impact abundances. The KIN inner spatial regime spectral features were keyed to {\it Spitzer} spectral features in the 4/16 and 4/26 spectra.  Similarly, the KIN outer spatial regime, light predominantly from a region $\sim$ 17 AU from the WD, is keyed to the {\it Spitzer} spectrum taken on 9/9 \citep{eva07c} as it is assumed that the abundances of atomic species in the latter, nebular {\it Spitzer} spectrum, would reasonably be representative of the environment observed by that KIN channel.  

Note that other symbiotic novae (e.g. V1016 Cyg, RR Tel) have broad emission features  around 10 and 18 $\mu$m, evident in our {\it Spitzer} spectrum, that have been successfully fitted by crystalline silicate features (e.g. \citet{sac07, sch01, eyr98}).  Motivated by this fact, we calculated the temperature and emission SED of various species of dust at the range from the pseudo-photosphere of the WD at which the KIN's outer spatial regime has maximum transmission. The center of the first constructive fringe, when the null fringe is located on top of the WD, at  9.8 $\mu$m and a projected baseline of 84 m, is $\sim$ 12 mas from the WD.  At a distance to the object of 1.4  kpc \citep{bar08} this corresponds to about 17 AU.   At day 3.8,   the pseudo-photosphere of the WD has a luminosity of 1.6$\times 10^5$ $L_{\odot}$ (radius, 18.1 $R_{\odot}$, and
temperature, 27050 K), computed by 
interpolating between values for the post-outburst evolution displayed in Table 5. 
The temperature of a black dust grain in thermal equilibrium is 272 K, using Eqn. A3  from
the Appendices in this paper, and is well below the sublimation temperature of silicate dust, $\sim$ 1500 K.
The dust feature in the {\it Spitzer} 9/9 spectrum is wide enough at FWHM that it falls across over seven spectroscopic elements in the KIN outer spatial channel meaning that the dust is both spectrally resolved and spatially localized.  

\begin{figure}
\begin{center}
\includegraphics[width=4.5in]{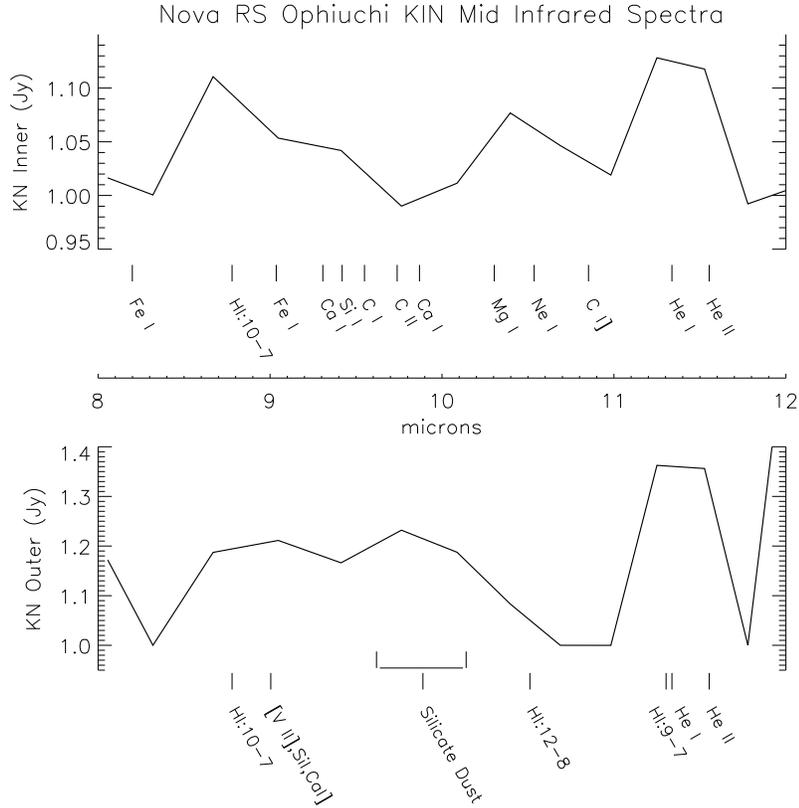}
\caption{Continuum normalized \textit{Keck} Interferometer Nuller spectra 3.8 days after peak V-band brightness.  Spectra are shown matched with lines identified in \textit{Spitzer} spectra.  The inner spatial regime line identification was matched to the cosmic abundances seen in the earlier \textit{Spitzer} spectra while the outer spatial regime was biased towards the later spectrum on September 9, 2006 that included a solid state feature and were assumed to have primarily nebular abundances. Please see the text for more discussion of the line identification process. \label{keckspitzlines}}
\end{center}
\end{figure}

Figure \ref{keckspitzlines} displays the association of identified {\it Spitzer} atomic emission lines with the continuum-normalized KIN inner and outer regime spectral features. Clearly the two traces are markedly different.  Note that when the inner and outer spatial regimes are summed, the result closely follows the re-binned {\it Spitzer} spectra with the exception of the KIN outer spatial regime feature (lower trace) centered at 9.8 $\mu$m.  Based on our assumptions of a primarily nebular environment in the vicinity of the outer spectrum, the atomic metals evident in the upper trace (KIN Inner spatial regime) would be unlikely to contribute to this feature.  In any case, the total power in these metal lines in the wavelength range of this feature (cf. Table \ref{spitzlinelist}), including Ca I, Si I, C I, \& C II, evident in the upper trace is 0.05 Jy, while that in the broad feature centered at 9.8$\mu$m in the lower trace exhibits is 0.24 Jy.  Our models suggest that the source of this feature may be hot silicate dust in the temperature range 800 - 1000 K. 

Note that our KIN data detects the faint emission from  well outside of the blast radius, assuming an initial shock front velocity of 3500 km/s \citep{obr06} and negligible deceleration.  The radiation from
this spatial region originates primarily from material around the nova that has been illuminated and warmed by photons from the nova flash and, as a result, must have existed {\it before} the nova event.  This establishes  that silicate dust, created in the vicinity of the RS Ophiuchi system some time previous to the 2006 outburst, is detected by our measurements, and is consistent with the conclusions of \citet{eva07c}.

\section{A physical model of the recurrent nova}

One aspect of the RS Oph binary system that has been neglected in the current literature
regarding the recent outburst 
is the effect of the motion of the two stars through the wind created by mass loss from the red giant
 star.  Garcia (1986) suggested that there could be a possible ring of material of
 diameter less than 40 AU around the RS Oph binary system or possibly surrounding the red giant
 component, based on his measurements of an absorption feature in the core of the Fe II emission
 line profile  at 5197 \AA.  The observations were performed in 1982 and 1983, several
 years before the 1985 nova outburst. 
 
Subsequently, and motivated by somewhat different observations, 
\citet{mas99} computed three-dimensional hydrodynamical 
models of morphologies of the envelopes of binaries with detached WD and RG/AGB components in general.  Their purpose was to see if these models could
reproduce some of the observed characteristics of axisymmetric or bipolar pre-planetary
nebulae.    Their study focused on a parameter space that encompassed outflow velocities
from 10 to 26 km/s, circular orbits with binary separations from 3.6 to 50 AU, and binary
companions having a mass range of 0.25  to 2 M$_{\odot}$.    For binary separations of
about 3.6 AU and mass ratios of 1.5, it was possible to generate a single spiral shock that
winds 2-3 times around the binary before it dissipates at $>$ 25 times the radius of the
RG star.  There is a density enhancement of about a factor of 100 over the normal density
in the wind in the plane of the orbit of the two stars, and an under density or evacuated region
perpendicular to the plane of the orbit.   Observational support for this model was found recently 
by \citet{bod07} who detected a double-ring
   structure in HST data, which they interpreted as due to an equatorial density
   enhancement and \citet{mau06} who observed a spiral pattern around the AGB star AFGL 3068 both of which were consistent with the model of \citeauthor{mas99}.  The underlying binary, a red giant and white dwarf,  was discovered by \citet{mor06}, who also determined the binary separation and 
hence approximate orbital period, which was consistent with expectations from the appearance
of the spiral nebula pattern seen by \citeauthor{mau06} and the model.

Figure \ref{dust-sub_01} shows the proposed geometry of the nebula in the plane of the orbit of the RS Oph system based
on the parameters adopted from \citet{dob94} for the system, including an orbital period of 460 days,
and an inclination angle of about 33 degrees, for the epoch just prior to the nova outburst.
The spiral shock model produces an
archimedian spiral nebula, with the separation between adjacent windings of about 3.3 mas
based on the period noted above and a wind speed from the red giant star of 20 km/s.  Based on 
these parameters
we estimate that about 17 such rings could be created between outbursts, with the overall size
of the nebula of the order of 100 mas.   However,
there could be as few as 10 rings to as many as about 20 rings, depending on the parameters, some
of which are not well known. 

In order to better understand the Keck Nuller observations and other high angular resolution
observations, we have modeled the evolution of the circumstellar nebula following the outburst,
which we display in Fig.  \ref{CSMatterEvol}.  For the purposes of this discussion, we adopt the model of 
Hachisu and Kato (2001),
 but with the simplification of a flat accretion disk geometry, i.e.,  not warped as in their paper.   
The details of our calculations are presented in the Appendices to this paper.  

 The results are of fundamental importance to the interpretation of our KIN data.   We begin
 with the most immediate effect after the blast, which is the sublimation of dust within a zone
 where the temperature of a blackbody dust grain would be $>$ 1500 K in equilibrium.  Figure \ref{dust-sub_02}
 displays the evolution of the dust sublimation radius, which is roughly 4-5 mas until about 
 day 70, after which it steadily declines to $<<$ 1 mas about 250 days after the outburst. 
 This means that much or all of the dust within this zone has entered the gaseous phase, except
 for a small ``sliver'' of material in the shadow of the red giant star.    This provides additional
 hot gas (rich with metals)  that is subsequently affected by the blast wave passing through 
 within the next few days.    
 
 As the two stars move relative to each other in their orbit about their common center of mass, 
 the location of the shadow of the nova moves, and consequently the material in the shadow that
 has not been affected by the blast wave from the nova  will be sublimated during this luminous
 phase, creating hot gas in the vicinity within a few mas of the stars.  This material may still have
 the type of repeating density structure that was initially present, and may be observable 
 with high angular resolution instrumentation. 
 
 Whether or not  this particular material is observable,  depends in part 
 on the density distribution as a function of latitude of the plane of the orbit of the two stars.
 If the material is uniformly distributed over 4$\pi$ steradians, then the fraction
 of the total solid angle subtended by the red giant star as seen by the white dwarf star is 
 approximately 0.13-0.29\% for a RG star between 27 and 40 R$_{\odot}$ and a binary separation of 
 1.72 AU.  However, the hydrodynamical studies of Mastrodemos and Morris (1999) show that
 the density falls off steeply as a function of latitude, and is as low as 1\% of the mid-plane density by
 latitudes of about $\pm$ 40 degrees.  This gives a scale height of about 9 degrees.  Thus, the
red giant star subtends a much bigger fraction of the solid angle up to this scale height,
i.e.,  to as much as 1.9\%.  

Furthermore, studies of the supernova blast waves around red giant stars indicate that the 
blast wave diffracts around the RG star, with a hole in the debris of angular size $\sim$ 31-34 degrees,
which did not depend strongly on whether the companion star was indeed a RG star or 
a main sequence or  subgiant star (Marietta, Burrows, \& Fryxell 2000).   In this case the fraction
of the sky subtended by this hole in the debris field is $\sim$ 10-13 \%.  If the material is concentrated
in the mid-plane, the effect is substantially bigger as noted in the previous paragraph.  Thus 
there are several reasons to expect that material in the shadow of the RG star can have an observable
effect.   

Stripping of material from the  red giant companion by the blast wave for supernova has been studied by Wheeler, Lecar, and McKee (1975) and by Lyne, Tuchman, and Wheeler (1992), however, Lane et al. (2007) showed  that stripping of material for the RG star is negligible for the less energetic blasts from RS Oph. 

 Another effect is the heating of the surface of the red giant star
 that faces the nova (see Appendices), as seen in Figure \ref{RG-face-temp}, for times past the maximum in the visible light curve.  Our calculations indicate that this side of the red giant star increases from about 
 3400 K to as much as 4200 K within a few days past the outburst.  The temperature steadily declines
 from day 70 until it reaches equilibrium with the other side around day 250.   The surface of the 
 RG star is initially heated by the shock from the blast wave (not included in our calculations), however, the continued heating  during the
 high luminosity phase is important as it affects the interpretation of data from modern complex
 instrumentation such
 as stellar interferometers and adaptive optics systems, where various subsystems are controlled at wavelengths  other than the measurement wavelength.
 
 
Figure  \ref{CSMatterEvol}  displays a schematic view of the system geometry  from days
4 to 90 after the recent outburst.  Assuming typical wind velocities of about 20 km/s for the red giant wind,
there are roughly 17 rings separated by approximately 3.3 mas that form between RS Oph outbursts.
The top left panel displays the system geometry at 4 days post-outburst.  A gray ring is drawn in the
center of the figure to indicate the size of the region affected by the blast at this epoch.  
In (a) the outer part of the spiral
is overlaid with light gray to indicate that it is not known if the material stays in a coherent spiral
past the first few turns.  The diameter of the shocked region is about 8.8 mas assuming the
blast wave travels at a velocity of $\le 1800$ km/s {\it in the plane of the orbit}, using the velocity measured
 by \citet{che07}, at approximately the same epoch as our measurements. (We acknowledge that most out-of-plane blast-wave velocities noted in the literature are much higher than this.) 
In this figure we assume the blast wave moves at constant velocity as it traverses the 
spiral shock material.  We show the extreme case in which
the blast wave is 100\% efficient in sweeping up material, thus creating a ring-like structure that propagates outward from the
system.  Note this figure is meant only to be illustrative and differs in detail from estimates of the
position of the blast wave from observations, such as \citet{obr06} who obtained a value
of the shock radius of 8.6 mas at day 13.8 by which time the blast wave had apparently slowed considerably.  We obtain a value of 10.5  mas for the shock radius assuming a constant velocity of $\sim$ 1800 km/s from one day past the initiation of the TNR process,
which we take to be about  3 days before maximum light in V band \citep{sta85}.   The value of 8.6 mas at day 13.8 is 
consistent with a somewhat lower mean velocity, of the order of 1400 km/s.
It is beyond the scope of this paper to compute the evolution of the
blast wave, however, this figure makes a connection with the previous observational evidence
for a ``ring" of material concentrated in the plane of the orbit as discussed by Garcia  (1986) and with 
the recent observations of \citet{che07} who observe different velocities perpendicular
to the plane of the orbit than in the plane of the orbit.  

The evolution of the luminosity of the red giant and nova is also significant  in aiding our
understanding of the observations.  Figure \ref{LumEvol}  (a)  displays computed V band light curves 
up to day 250 after the outburst  including the effect of the irradiation of the 
accretion disk and  the irradiation of the red giant star by the nova.  Note this calculation overestimates
the total luminosity of the nova  during the
period from about 10 days to about 50 days, and this is likely due to the simplified disk 
geometry that we have employed in our own calculations.  Most importantly these 
light curves show that the V band luminosity is dominated by the nova for about the first 50-70 days, and after that the red giant star dominates the V band luminosity.  This is significant as most telescopes
track on V band light (including interferometers) and the tracking center then moves by a
mas or so during the post-outburst evolution of the system.  H band luminosity evolution
is plotted in Fig. \ref{LumEvol} (b), and is different than that of the V band evolution.  At H band the nova 
dominates the luminosity only for the first 5 days or so, and after that the H band light curve is
completely dominated by the red giant star.  This means the phase center for fringe detection 
is offset from that of the tracking center from days 5 onward by a mas or so, as mentioned above.
The N band luminosity, displayed in Fig. \ref{LumEvol} (c), 
 evolves  like that of the  H band, and the nova dominates the mid-infrared
light only up to day 4.  After that there is an offset between the tracking center and N band 
fringe center like that noted for the H band. 

\section{Implications to the Keck Nuller  and other high-angular resolution observations}

We now reexamine  data from the 2006 outburst
within the framework of the spiral shock model of the geometry of RS Oph presented in the last section.
In its simplest form, this geometry evolves into a bipolar morphology similar to that seen
in the radio emission from the outburst that was discussed by O'Brien et al. (2006), as seen
in Figure \ref{CSMatterEvol}(d).   

Such a geometry provides a natural explanation for some of the differences between the 
interferometric and other measurements, as the blast wave would be restricted and slowed in the orbital plane due to 
the high density regions, while its flow would be relatively unimpeded  perpendicular to
that plane.  This corresponds well to what was observed by Chesneau et al. (2006),
where the observed velocity components came from distinctly
different position angles.  These authors also calculate that the red giant star should be at position
angle 150-170 degrees at the time of the outburst. Recently, \citet{bra08} have derived spectroscopic orbits for both components of the RS Oph system based on the radial velocities of the M giant absorption lines and the broad emission wings of H$\alpha$. They have also constrained the orbit inclination, $i \geq 49 \pm 3$, using the estimated hot component mass, $M_{\rm h}\, \sin^3 i = 0.59\, \rm M_{\odot}$, and assuming  that the white dwarf mass cannot exceed the Chandrasekhar limit.  Thus the plane
of the orbit is such that the high velocity flow would be expected to be mostly East-West and 
the the continuum emission would have an elliptical shape with the position angle measured
by Chesneau et al. (2006).   Furthermore, the spiral shock wave tends to have the largest
densities near the white dwarf star and on the opposite side of the red giant star, the separation
being of the order of two to four times  the separation between the two stars, i.e., a few 
mas.  The larger separation occurs when
the red giant star has been spun up so that it is synchronous with the orbital 
motion.  Thus it is possible that some of the data can be interpreted in terms of emission
from hot clumps of material within the spiral shocks.     Detailed radiative transfer calculations 
will need to be performed to refine this model.  

\begin{figure}
\begin{center}
\includegraphics[width=4.5in]{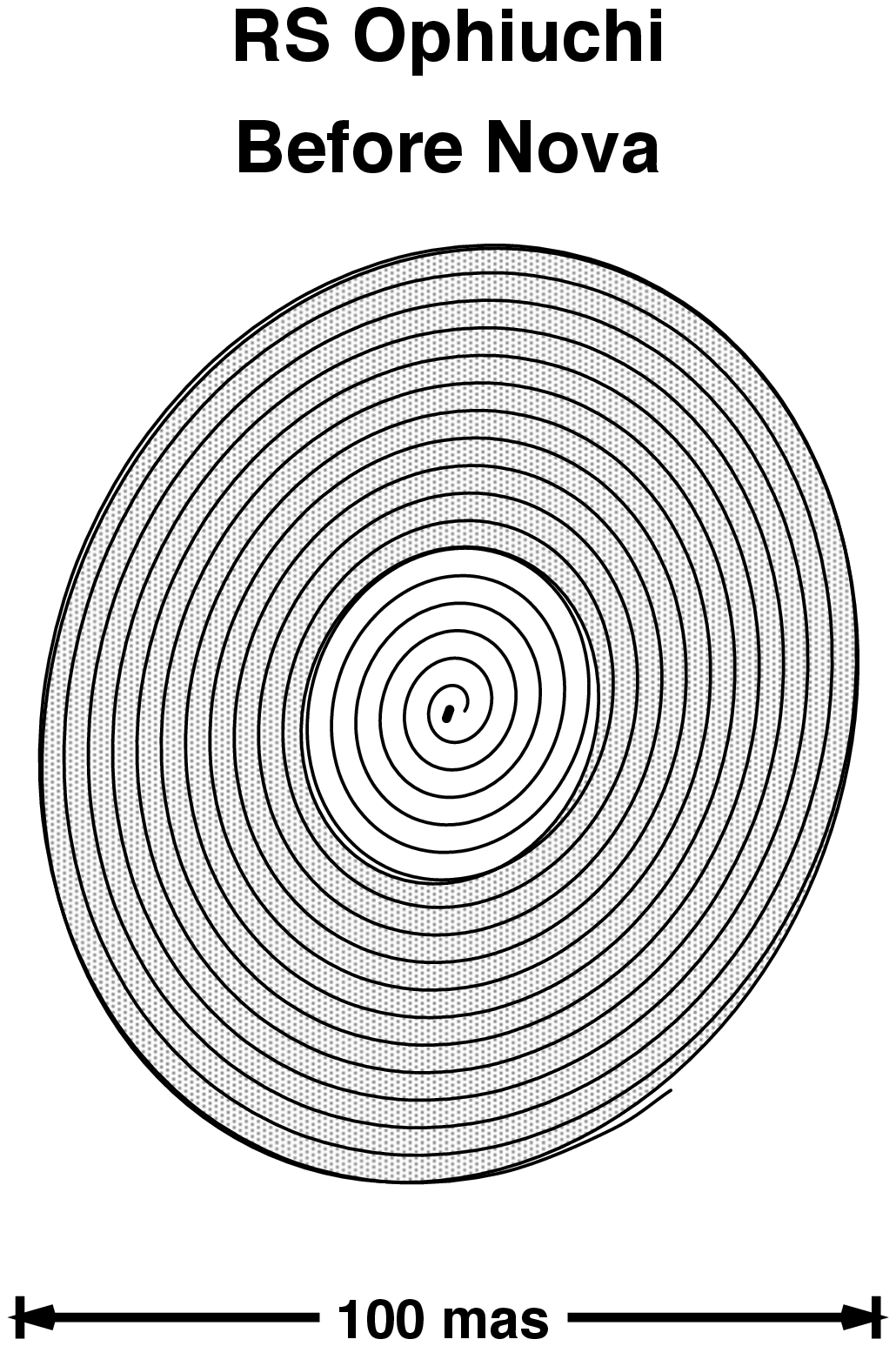}
\caption{Proposed model of circumstellar material surrounding the binary RS Oph before the nova
eruption.  The interaction of the white dwarf and red giant star in the slow dense wind of the red giant
star can create a spiral shock with an enhanced density in the plane of the orbit of the two stars.  The overall size of the dense in-plane material is of the order of 100 mas if RS Oph is at a distance of 1400 pc and the wind speed is about 20 km/s.  \label{dust-sub_01}}
\end{center}
\end{figure}

\begin{figure}
\begin{center}
\includegraphics[width=4.5in]{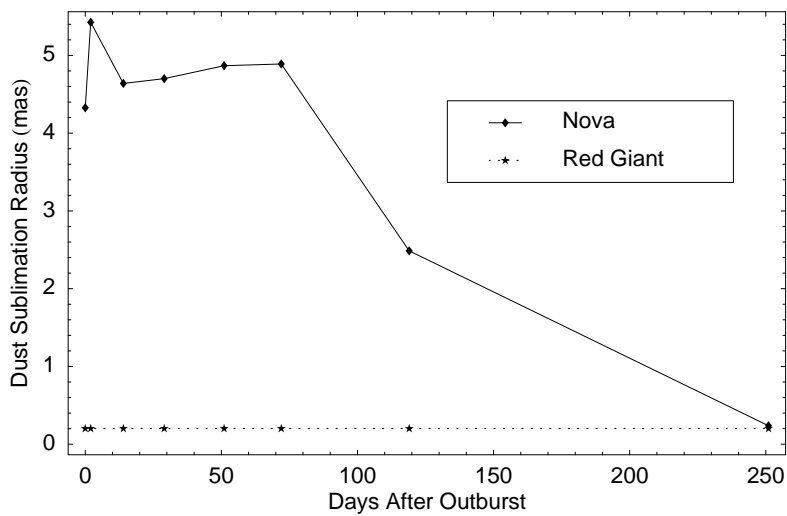}
\caption{Sublimation radius of dust (mas) as a function of the number of days post-outburst for the Nova
and for the red giant companion.  \label{dust-sub_02}}
\end{center}
\end{figure}

There is a wealth of recent observational work on RS Oph across the spectrum from X-ray \citep{sok06, bod06},
near-infrared \citep{eva07a, che07, das06},  radio \citep{obr06}, and now mid-infrared \citep[this work]{eva07b, eva07c}.  
Generally speaking, the observational picture is consistent with the shocked wind model
as described by \citet{bod85}.  Extensive modeling of the light curves has been 
performed by \citet{hac00, hac01}, which include effects such as the irradiation of the
red giant star by the white dwarf, and the
accretion disk, which become important
about 4 days post-outburst, and which are the basis for 
our own calculations presented in Figs. \ref{dust-sub_01},\ref{dust-sub_02}, and \ref{LumEvol}.  In this general picture the recurrent nova 
differs from a classical nova in that the high-velocity  ejecta from the WD is impeded by the 
wind from the red giant star, which in turn generates a shock wave that propagates through the
red giant wind.   Observations of the shock wave in the near-infrared
by \citeauthor{das06} confirm this picture and 
trace the evolution of  the widths of Pa $\beta$ and O I  lines
as a function of time post-outburst, and clearly show
the expected free-expansion phase of the shock ended about 4 days post-outburst.  This is also consistent with independent measurements conducted by \citet{bod06} in which this phase was proposed to have ended by approximately day 6. 

Most of the observational work has used spectroscopic methods,
but only the interferometric observations
are capable of spatially separating the various components of RS Oph that contribute to the 
emission that is seen spectroscopically.  Chesneau et al. (2007) observed RS Oph  5.5 days post-outburst in the 
continuum at 2.13 $\mu$m,  Br $\gamma$ at 2.17 $\mu$m, and He I at 2.06 $\mu$m with the 
AMBER instrument on the VLTI.  They fitted their data with 
uniform ellipses, gaussian ellipses, and a uniform ring.  The models had excellent consistency
in terms of the position angle of the ellipse ($\sim$140 degrees) and the ratio between minor and 
major axes (0.6).  For the continuum at 2.13 $\mu$m, a uniform ellipse had a major axis of
4.9 $\pm$ 0.4 and minor axis 3.0 $\pm$ 0.3 mas, while the gaussian ellipse had a major 
axis of 3.1 $\pm$ 0.2 and minor axis of 1.9 $\pm$ 0.3 (FWHM).   These measurements
are consistent with the expected size of the shocked region at the epoch of their measurements.

Monnier et al. (2006) presented IOTA results in the near-infrared bands at H and K,
and had somewhat different conclusions 
than some of the other workers because they were able to fit their visibility and closure phase
data best with a binary model with two sources separated  by  3.13 $\pm$ 0.12 mas, position
angle of 36 $\pm$ 10 degrees, and brightness ratio of 0.42 $\pm$ 0.06.  Their gaussian 
fits at 2.2 $\mu$m gave a FWHM of 2.56 $\pm$ 0.24 mas for the same period of time as the AMBER observations.
A striking feature of those results is that the size of the emitting region 
decreased 10-20\% between about days 4 and 65 post-outburst.  For example, the size
at 2.2 $\mu$m actually decreased from about 2.6 to 2.0 mas (FWHM), while at 1.65 $\mu$m, the
size decreased from 3.3 to 2.9 mas.   However the 2 $\mu$m 
continuum sizes are in approximate agreement.   Monnier et al. (2006) rule out an expanding fireball
model, however, they would have over-resolved the fireball anyway since it would be about  8.8 mas in diameter, at a distance of 1600 pc, or a substantially larger angular size  if the distance
were smaller.   Lane et al. (2007) observed RS Oph also with IOTA, PTI, and the Keck Interferometer
at H band over a longer time period, up to 120 days from the V band maximum.  They observed
an increase in diameter from approximately 3 to 4 mas from days 0 to 20 and a decrease to 
less than 2 mas in diameter around day 120.  They interpreted the near-infrared size data in terms 
of a simplifed model of  free-free emission in the postnova wind as the 
mass ejected in the wind decreased during that time period.

The  emitting region at 10 $\mu$m is somewhat
larger than that seen at  2 $\mu$m, for example,  our data are fitted by a gaussian  4.0 $\pm$ 0.4 mas
(FWHM), while the IOTA data had a size of 2.6 $\pm$ 0.2 mas.  By 4 days post-outburst, a spherical
shock wave would be expected to have a radius of about 3.9 mas assuming a speed of approximately 1730  km/s (O'Brien et al. 2006)
and a distance of 1.4 kpc.    Thus, according to this simple model, all the interferometric 
IR continuum emission should be coming from within the post-shock region.  The AMBER/VLTI
results indicate a more complex picture of the velocity field of the expanding material, with 
two indicated - a ``slow'' field between -1800 and 1800 km/s, and a ``fast'' one between 
-3000/-1800 km/s and 1800/3000 km/s.    The position angle of the emitting material for the two
velocity groups differ, with  the ``fast'' component being well defined in the East-West (position
angle 90 or 270 degrees $\pm$ 5 degrees) direction
and the ``slow'' component with position angles  from 55 to 110 degrees modulo 180 degrees.

Recent observations by Bode et al. (2007) using the HST confirm the elliptical shape seen with
AMBER, with an axial ratio of 0.5-0.6.    The HST observations also confirm the velocity structure 
seen by AMBER, i.e., the ``fast'' velocity field in the East-West direction, of the order of 3000 km/s,
and the ``slow'' velocity about half that of the ``fast'' one.  The HST results showed that the expansion
rate in the plane of orbit of the stars decreased, from about 0.62 mas/day on day 13.8 to 0.48 mas/day
on day 155.  This is a reduction in velocity from about 1700 km/s to about 1300 km/s.

The spiral shock model is clearly relevant to the interpretation of the IOTA data presented in Monnier 
et al. (2006).   As noted above, the increased density of gas and dust in the arms of the spiral
would certainly provide an impediment to the free expansion of the fireball, as well as provide
a reservoir of hot material that would emit strongly at 2 $\mu$m.  Their data indicate a closure 
phase that is consistent with zero or 180 degrees for the first epoch (days 4-11) and convincingly
non-zero only for hour angles from -1.5 to -1 hr on the second epoch (days 14-29), and 
at +1/2 hr on the third epoch (days 49-65).  Hence, the closure phase signals and the binary 
interpretation may be more consistent with hot clumpy material that is cooling, some of which may
(re-) condense into dust
as the white dwarf's outburst luminosity declines from its most luminous state, i.e., as shown in 
Fig. \ref{dust-sub_01}.   Note that the sublimation radius  decreases from $\sim$ 5 mas to about 
2 mas from day 70 to day 120.  Furthermore, as the luminosity
changes, the star tracker on IOTA may be providing a different optical center to the fringe
detection system, since in the first few days during the outburst the optical emission is dominated
by that centered on the white dwarf star itself, whereas after it has cooled, the optical emission is dominated by
the red giant star.  Thus the effects of changes in the optical tracking and 
interferometric phase center will need to be included in the analysis
to properly interpret the data.  The offset between the optical tracking center and the
fringe phase center can cause a miscalibration of the visibility.
Since the 2 $\mu$m emitting region is actually
decreasing in size with time, as seen in their Fig. 1 and Table 2, it seems more likely they are observing 
the cooling of this hot material near the two stars than actually resolving the binary, however,
the effects of the material in the shadow of the red giant star must be included in the interpretation
of the near-infrared and mid-infrared data.


\section{Summary and Conclusions}
We  analyzed data from  the recurrent nova RS Oph for the epoch
at $\sim$ 4 days post-outburst using the new KIN instrument.
These data allowed us to determine the size of the emitting region around the RS Oph
at wavelengths from 8-12 $\mu$m.  
By fitting the unique KIN inner and outer spatial regime data, we  obtained an angular size of the mid-infrared continuum of 6.2, 4.0, or 5.4 mas for a disk profile, gaussian profile (FWHM), and shell profile respectively.  
The data show evidence of enhanced neutral atomic hydrogen emission located in the inner spatial regime relative to the outer regime.  There is also evidence of a 9.7 $\mu$m
silicate  feature seen outside of this region, which is consistent with dust that had condensed
prior to the outburst, and which has not yet been disturbed by the blast wave from the nova. 
Our analysis of the observations, including the new ones presented in this paper, are most 
consistent with a new physical model of RS Oph, in which spiral shock waves associated with
the motion of the two stars through the cool wind from the red giant create density enhancements
within the plane of their orbital motion.  

Further observations are needed to clarify this new picture of the RS Oph system.  One issue that
has not been fully resolved is whether or not the red giant star really overflows its Roche lobe.  
If so a hot spot would be expected where the material streaming from the red giant envelope 
hits the accretion disk, and UV or X-ray observations could search for this effect.
Another approach would be to observe  RS Oph over several orbital cycles using infrared photometry
to look for  variations of the light curve due to the departure of the red giant star from spherical
symmetry.  High resolution
spectra could also help.   Confirmation of the rotational velocity of the red giant
star measured by Zamanov et al. (2007) would be worthwhile, and could provide another 
estimate of the absolute size of the red giant star, assuming that it is co-rotating with the
orbit.  Further KIN observations within the next few years would also be helpful as they could
show evidence of the re-establishment of the spiral shock wave, and perhaps some information
about the shape of the circumstellar material and dust formation.  Another epoch of HST
observations would also determine the deceleration of the outflow in the two directions, i.e.,
within the plane of the orbit of the two stars and along the poles.  

Theoretical studies of the motion of the blast wave in an environment with a high density 
region in the plane of the orbit of the two stars are also worthwhile.  In particular, it would be
important to understand how the blast wave is diffracted around the red giant star and 
how it propagates in an medium with the periodic density enhancements in the plane due to
the spiral pattern.  

The recurrent nova RS Ophiuchi is a rich system for the study of circumstellar matter under
extreme physical conditions.  Continued study will provide important insights into Type Ia
supernovae, of which RS Oph may be a progenitor.

\acknowledgments

We are grateful to the National Aeronautics and Space Administration, Jet Propulsion Laboratory, the California Association for Research in Astronomy, the Harvard-Smithsonian Center for Astrophysics
(including SAO grant G06-7022A to JLS), and to the National Aeronautics and Space Administration, Goddard Space Flight Center for support of this research.    The
data presented herein were obtained at the
W.M. Keck Observatory, which is operated
as a scientific partnership among the California
Institute of Technology, the University
of California and the National Aeronautics
and Space Administration. The Observatory
was made possible by the generous financial
support of the W.M. Keck Foundation.  
JPW acknowledges support provided by an NPP Fellowship (NNH06CC03B) at NASA Goddard
Space Flight Center. This work has made use of services produced by the Michelson Science
Center at the California Institute of Technology.  One of the authors (RKB) would also like to acknowledge Eugene E. Rudd formerly of the United States Naval Research Laboratory for his continued support and encouragement.   The authors thank the referee, Dr. Michael Bode, for his careful and thorough review of the manuscript, which has helped us to  significantly improve it.

Facilities: \facility{Spitzer}, \facility{IOTA}, \facility{Keck:I}, \facility{Keck:II}.

\appendix
\section{Luminosity Evolution}
In this appendix we discuss the mathematical formulation used to derive the evolution of
the luminosity of the white dwarf star starting at the maximum magnitude in V band.  Our 
calculations are based on those presented in the paper by Hachisu and Kato (2001), who
expanded on their discussion of their light-curve model presented in Hachisu and Kato (2000).  

Our purpose is to evaluate the light curve not only at V band, but also at H, K, and N bands,
which have been used by the AMBER, IOTA,  Keck, and PTI  interferometers to observe RS Oph
after the 2006 outburst.  Our discussion is in fact relevant to any modern instrument that has 
wavefront sensing and control, and also tracking, at wavelengths different than that of the 
observations of interest.  For example, the IOTA interferometer performs its precision pointing at
V band, but observations are made at H band.  Similarly the Keck Interferometer does 
precision pointing at V band, wavefront control of the two large apertures at H band, and 
the data are taken at N band.  This difference in wavelengths is important because as
the V band luminosity of the white dwarf star decreases, the observed light from the binary
is mainly from the red giant star, which gives a shift in the center of light.  This shift can affect 
the calibration of these instruments as there is an offset, which will change over time as compared
to a calibrator, which has all the same optical center at all times for all of these wavebands.  

The evolution of the stellar parameters for the white dwarf are given in Table \ref{evo} of  Hachisu
\& Kato (2001). 
We assume the parameters for the red giant star remain constant with $R_{rg} = 40 R_{\odot}$,
and $T_{rg}$ = 3400 K.  

The blackbody flux density (erg cm$^{-2}$ s$^{-1}$ Hz$^{-1}$) at a frequency, $\nu$, for a star of temperature, T$_{\ast}$, radius, 
$R_{\ast}$, and distance, $D$ is:
\begin{equation}
F_{\ast}(\nu) = 2 \pi (R_{\ast} / D)^2  (h { \nu}^3 / c^2) / [exp(h \nu / k T_{\ast}) - 1] ,
\end{equation}
where $h$, $k$, and $c$, are Planck's constant, the Boltzmann constant, and the speed of 
light, respectively.

The total luminosity (erg s$^{-1}$) is:
\begin{equation}
L_{\ast} = 4 \pi \sigma  {R_{\ast}}^2 {T_{\ast}}^4 ,
\end{equation}
where $\sigma$ is Stefan's constant.

\subsection{Sublimation Radius}
The temperature of a  black grain in radiative equilibrium at a distance, $r$, 
from a star with luminosity, $L_{\ast}$ is:
\begin{equation}
T_{grain} = \left(\frac{L_{\ast}}{16 \pi \sigma r^2}\right)^{1/4}  = 
\frac{T_{\ast}}{\sqrt{2}}  ~ \left( \frac{R_{\ast}}{r}\right)^{1/2}
\end{equation}
A simple rearrangement of the above equation yields a formula for the sublimation radius, $R_{sub}$,
assuming a sublimation temperature, $T_{sub}$:
\begin{equation}
R_{sub} = \left(\frac{L_{\ast}}{16 \pi \sigma T_{sub}^4}\right)^{1/2} 
 = \frac{R_{\ast}}{2}   ~ \left(\frac{T_{\ast}}{T_{sub}}\right)^2 .
\end{equation}
For this paper we assume a sublimation temperature of 1500 K.

The evolution of the  sublimation radius for the red giant star (dashed lines) and nova (solid lines) beginning at the maximum in V band is
displayed in Fig. \ref{dust-sub_02}.  Note that the sublimation radius remains approximately constant at about 5 mas for the first 70 days, after which it gradually reduces to $<$ 0.2 mas after about day 120. 

\subsection{Illumination of the Red Giant Star}
During the high luminosity phase of the outburst, the surface of the red giant star facing the nova
is heated substantially.  We calculate the effect of the nova luminosity on the red giant star using
Eqn. (10) of Hachisu \& Kato (2001), which is based on thermal equilibrium between the 
faces of the two stars, which we display here:
\begin{equation}
\sigma {T_{rgi}}^4 =  \eta_{rg}  \cos ( \theta )  L_{wd} /  (4 \pi r^2)  + \sigma {T_{rg}}^4
\end{equation}
where $T_{rgi}$ is the new effective temperature of the red giant star for facing side and where 
we include an extra term from the white dwarf luminosity, $L_{wd}$.  There are two additional
constants included in this new term, the first is $\eta_{rg}$, which is an effective emissivity for 
the stellar surface, while the second, $\cos(\theta)$  is the average inclination angle of the surface.
The term, r, is the distance between the two stars, and $T_{rg}$ is the original temperature of
the red giant star.  Equation A5 can be rearranged into a simple form after substituting  for
$L_{wd}$: 
\begin{equation}
{T_{rgi}}^4 =  \left[ \eta_{rg}  \cos ( \theta ) \left( \frac{R_{wd}}{r} \right)^2 \right]  {T_{wd}}^4  + {T_{rg}}^4
\end{equation}
 Equation A5 was evaluated as a function of time, using the evolution of the 
white dwarf luminosity derived from Eq. A2 and the values from Table 5. 
The result of our computation of this effect is displayed in Fig. \ref{RG-face-temp}, in which we use $ r = 325 R_{\odot}$,
$\cos(\theta) = 0.5$, $\eta_{rg} = 0.5$, and $T_{rg} = 3400$ K. 

\subsection{Accretion Luminosity}

The computation of the evolution of the accretion luminosity is based on the treatment of 
Hachisu \& Kato (2001) with the simplification of a flat accretion disk, instead of adding
the complexity of a warped accretion disk that is used in their paper.  
These calculations are based on the well known \citet{Lynden74} and \citet{Pringle81}  equations for  the luminosity of an accretion disk, where the accretion luminosity is based on a numerical integration
of the flux from the accretion disk, assuming a temperature distribution across the disk.
Normally the temperature of the accretion disk is determined solely by the accretion rate onto
the disk for normal viscous heating of the disk.  To this Hachisu \& Kato (2001) added a
term based on additional heating of the disk from the white dwarf star as it is in a high
luminosity state and is contracting and heating during the constant high luminosity period 
after the outburst.   The flux density of the disk at frequency $\nu$ is given by:
%
 %
%
\begin{equation}
F_{disk}\left(\nu\right) =  \frac{2 \pi \cos(\theta_{inc})}{D^2} \int_{R_{in}}^{R_{out}}  \rho  B_{\nu}[T_{disk}(\rho)] d\rho
\end{equation}
The conventional expression for the radial distribution of the temperature of the disk is:
\begin{equation}
T_{disk1}\left(\rho\right) = \left(\frac{3GM_*\dot{M}}{8\pi\sigma \rho^3} \left[1-\left(\frac{R_*}{\rho}\right)^{1/2}\right]\right)^{1/4}
\end{equation}
where $T_{disk1}$ is the temperature in the accretion disk \citep{Pringle81} and D is the distance to the Earth. The quantity B$_{\nu}$ is the Planck function, M$_*$ is the mass of the white dwarf, R$_*$ is its radius, $\dot{M}$ is the accretion rate onto the disk, and $\sigma$ and G are the Stefan-Boltzmann constant and the gravitational constant, respectively. The maximum temperature in the accretion disk occurs at a radius R$_{max}$ = 1.36 R$_*$, and is given by
\begin{equation}
T_{max} = 0.488 \left(\frac{3GM_*\dot{M}}{8\pi \sigma R^3_*}\right)^{1/4}
\end{equation}
as discussed in \citet{Lynden74}, \citet{Pringle81}, and \citet{Hartmann98}. In this treatment we neglect the emission from the boundary layer, which occurs over a very small angular region around the surface of the white dwarf star and emits at a very high temperature, which would have a negligible contribution to the luminosity in the infrared region near 10 $\mu$m. 
The evolution of the luminosity of the accretion disk will be computed using the  formulation of
Hachisu and Kato (2001), as set up below.
%


      
 Let us now formulate the evolution of the accretion disk luminosity by adjusting some of the 
parameters in the equation above as a function of time.  In particular the outer radius of the
accretion disk 	$R_{disk}$ is parametrized by a power law decrease from day 7 until day 79
using the following formulae:
\begin{equation}
R_{disk} = \alpha {R_1}^{\ast}
\end{equation}

\begin{equation}
 \alpha = \alpha_0 { \left( \frac{\alpha_1}{\alpha_0} \right) }^{(t-t_0)/72}
\end{equation}
In these equations ${R_1}^{\ast}$ is the inner critical radius of the Roche lobe for the white
dwarf (nova) component, which we assume is 138.6 $R_{\odot}$
%
%
The parameter $\alpha$ helps define the size of the accretion disk which varies between $\alpha_0$ = 0.1 and $\alpha_1$ = 0.008, and so the power law form of Eqn. A10 is to make an interpolation function.   
Haschisu and Kato include irradiation of the accretion disk as the WD luminosity changes between days 7 and 79, where the accretion disk temperature is given by:
\begin{equation}
\sigma {\left[ T_{disk}(\rho)\right]}^4 = \eta_{disk} \cos(\theta_{inc}) \left( \frac{L_{wd}}{4 \pi \rho^2} \right) +
 \left(\frac{3GM_*\dot{M}}{8\pi  \rho^3} \left[1-\left(\frac{R_*}{\rho}\right)^{1/2}\right]\right)
 \end{equation}
%
%
The second term of this equation is the traditional accretion luminosity term, while the first term includes the luminosity of the white dwarf.  The parameter $\eta_{disk}$ is an efficiency factor, and is assumed to be 0.5, and $\cos(\theta_{inc})$ is 0.1, based on the average inclination of the surface.   They assume the outer radius of the accretion disk is at a temperature of 2000 K, and is not affected by radiation from the WD photosphere. 	 
For the remainder of the calculation we will compute $\alpha$ in discrete time intervals corresponding to what we have done before over the days that this effect matters.

The integral is numerically evaluated at the center wavelengths for V (0.55 $\mu$m), H (1.65 $\mu$m), and N (10.5 $\mu$m) bands,
using  an inner radius given by the radius of the white dwarf  star and the outer radius given by the
equation for $R_{disk}$, for the time steps of Table 5  and the white dwarf parameters in that
table.  The results are plotted in Fig. 10.  In this figure the quantities plotted include the the 
contributions of the red giant and white dwarf alone (stars and solid squares), the red giant
including effects of irradiation by the white dwarf (diamonds), the accretion disk alone (solid circles and
dash-dot-dot line), the irradiated accretion disk (triangles and dashed line), and the total for the
white dwarf including the irradiated accretion disk (diamonds and solid line).



\clearpage

\begin{deluxetable}{ccrrrr}
\tablewidth{0pt}
\tablecaption{Observing Log for RS Ophiuchi.\label{observations}}
\tablehead{
\colhead{Object} & \colhead{Type}& \colhead{Time}& \colhead{U}& \colhead{V}& \colhead{Airmass} \\
\colhead{} & \colhead{}& \colhead{(UT)}& \colhead{(m)}& \colhead{(m)}& \colhead{}
}
\startdata
Chi UMa & cal & $15\colon 07\colon 07$ &  23.85 & 80.61 & 1.38\\
Chi UMa & cal & $15\colon 15\colon 39$ &  21.92 & 81.25 & 1.41\\
RS Oph & trg & $15\colon 50\colon 15$ &  54.57 & 64.75 & 1.46\\
RS Oph & trg & $16\colon 03\colon 46$ &  55.35 & 64.37 & 1.39\\
RS Oph & trg & $16\colon 12\colon 35$ &  55.75 & 64.12 & 1.35\\
Rho Boo & cal & $16\colon 34\colon 24$ &  39.63 & 75.14 & 1.08\\
\enddata
\end{deluxetable}


\begin{deluxetable}{ccccc}
\tablewidth{0pt}
\tablecaption{RS Ophiuchi model fitting results.\label{models}}
\tablehead{
\colhead{Source Model} & \colhead{Angular Size (mas)}& \colhead{Radiant Flux} & \colhead{Major Size (mas) }& \colhead{Minor size\tablenotemark{a} (mas)} \\
\colhead{} & \colhead{(N band)}& \colhead{(Jy)} & \colhead{(K band)}& \colhead{(K band)}
}
\startdata
Uniform Disk & $6.2\pm0.6$& $22.4\pm3.9$ & $4.9\pm0.4$ & $3.0\pm0.3$\\
Uniform Gaussian\tablenotemark{b} & $4.0\pm0.4$ & $22.4\pm3.8$ & $3.1\pm0.2$  &$1.9\pm0.3 $ \\
Uniform Shell\tablenotemark{c} & $5.4\pm0.6$ & $22.4\pm3.8$ & $3.7\pm0.3$ & $1.9\pm0.2$\\
\enddata
\tablenotetext{a}{Sizes for continuum values at 2.3 $\mu$m after \citet{che07}.}
\tablenotetext{b}{Full width at half maximum.}
\tablenotetext{c}{Spherical shell with thickness 1.0 mas - optically thin.}

\end{deluxetable}

\begin{deluxetable}{cccccc}
\tablewidth{0pt}
\tablecaption{Mid Infrared \textit{Spitzer} Line List: N-band.\label{spitzlinelist}}
\tablehead{
\colhead{Wavelength} & \colhead{ID} & \colhead{{\it Spitzer} 4/16} & \colhead{{\it Spitzer} 4/26}  & \colhead{\it Spitzer 9/9} \\
\colhead{($\mu$m)} & \colhead{} & \colhead{(Jy)} & \colhead{(Jy)}  & \colhead{(Jy)} 
}
\startdata
7.460 & $Pf\alpha - HI \colon 6-5$& detected \tablenotemark{a}&   detected & detected\\
7.652 & $[NeVI]$ & detected&  detected & detected\\
8.180 & $FeI$ & 0.009 & 0.007 &--\\
8.760 & $HI \colon 10-7$ & 0.042&  0.038 & 0.020 \tablenotemark{b}  \\
8.985 & $[V II]$, Si I, \& Ca I]& -- & -- & 0.010\\
9.017 & $FeI$ & 0.015 &0.021&--\\
9.288 & $CaI$&0.004 &0.004&--\\
9.407 & $SiI$ & 0.014 &  0.013 & -- \\
9.529 & $CI$&0.009 &0.008&--\\
9.720 &$CII$& 0.018&0.017&--\\
9.852 &$CaI$ &0.007&0.008&--\\
10.285 &$MgI$ &0.007&0.006&--\\
10.492&$HI \colon 12-8$&--&--&0.013\\
10.517 &$ NeI$ &0.026&0.026&--\\
10.833 &$ CI]$ &0.018&0.016&--\\
11.284&$HI \colon 9-7$&--&--&0.007\\
11.318 &$ HeI$ &0.060&0.053&--\\
11.535 &$ HeII$ &0.016&0.013&--\\
12.168 &$HI \colon 36-11$&0.011&0.009&--\\
12.372 & $Hu\alpha - HI \colon 7-6$ & 0.158 &  0.150 & 0.037  \\
12.557 &$SiI$ &0.040&0.036&--\\
12.803 & $[NeII]$&--&--&0.159\\
12.824 &$HeI$ &0.009&0.015&--\\
13.128  &$HeII$& 0.009&0.011&--\\
13.188  &$HI \colon 18-10$ &0.009&0.010&0.005\\

\enddata

\tablenotetext{a}{Detected in \textit{Spitzer} spectrum but not fit due to intrusion of band edge.}
\tablenotetext{b}{Blended with neutral hydrogen lines at 8.721 and 8.665 $\mu$m }
\tablecomments{Some of these species were blended with others, and, due to the difficulty in deblending Spitzer low-resolution channel spectra, may in some cases be misidentified. This is not critical, however, to the conclusions of this paper.}

\end{deluxetable}
\clearpage
\begin{deluxetable}{cccccc}
\tablewidth{0pt}
\tablecolumns{4}
\tablecaption{Continuum-normalized, Mid-infrared KIN Emission Source Identification.\label{kinfeaturelist}}
\tablehead{

\colhead{Center Wavelength} & \colhead{Spectroscopic Width} & \colhead{Flux} & Attributed to\\
\colhead{($\mu$m)} & \colhead{($\mu$m, FWZM)} & \colhead{(Jy)} 
}
\startdata

\cutinhead{KIN Inner Spatial Regime}
8.7 & 8.3 - 9.1\tablenotemark{b}& 0.06\tablenotemark{b} & H I: 10-7, FeI, Ca I\tablenotemark{c} \\
9.4 &  8.9 - 11.1 & 0.02 & Ca I, Si I, C I, C II\\
10.4 & 9.9 -  11.1 & 0.04 & Mg I, Ne I, C I]\\
11.4 & 11.1 -  11.8 & 0.07 & He I, He II\\
12.5 & 12.1 - 12.6 & 0.15 & Hu$\alpha$\\

\cutinhead{KIN Outer Spatial Regime}

8.9 & 8.3 - 9.5 & 0.14& H I: 10-7, [V II], Si I, Ca I]\tablenotemark{d}\\
9.8 & 9.0 - 10.7 &  0.24&Silicate Dust\tablenotemark{e} \\
11.4 & 11.0 - 11.8 & 0.19 & H I: 9 - 7, He I, He II\\
\enddata

\tablenotetext{a}{Approximate FWZI continuum crossing points.}
\tablenotetext{b}{Because \textit{Keck} and \textit{Spitzer} data were not taken simultaneously and because of the extreme transient nature of the RN outburst it is not possible to assert particular flux numbers to the atomic lines attributed to the KIN feature.  The flux listed should be considered the maximum, bounding flux for any one of the atomic species shown here. }
\tablenotetext{c}{All atomic line emission in the KIN inner spatial regime is assumed to be emitted by a species at approximate cosmic abundance as for RN only a very moderate amount of nucleosynthesis is theorized to occur.}
\tablenotetext{d}{All atomic line emission in the KIN Outer spatial regime are assumed to be predominantly of nebular abundance with some contribution from uncondensed metals.}
\tablenotetext{e}{Spectral feature is spectrally and at least partially spatially resolved. Silicate dust feature has more than seven spectral elements across it and emits relatively strongly at a distance centered $\sim$ 17 AU from the WD. All flux is assumed to be attributable to emission by the dust.}

\end{deluxetable}

\begin{deluxetable}{l@{\extracolsep{10ex}}ll}
\tablewidth{0pt}
\tablecaption{Post-outburst Evolution of the White Dwarf Star in RS Oph \tablenotemark{a}\label{evo}}
\tablehead{
\colhead{Time\tablenotemark{b}} & \colhead{Radius}& {Temperature} \\
\colhead{(days)} & \colhead{(R$_{\odot}$)}& {(K)} \\
}
\startdata
0 & 45 & 12200 \\
2 & 21 & 20000\\
14 & 1.6 & 67000 \\
29 & 0.56 & 114000 \\
51 & 0.23 & 181000 \\
72 & 0.083 & 302000\\
119 & 0.0037 & 1020000 \\
251 & 0.003 & 350000 \\
\enddata
\tablenotetext{a}{Stellar parameters for the white dwarf star after the 2006 outburst
of RS Oph based on Hachisu \& Kato 2001}
\tablenotetext{b}{The zero time for these parameters is the V band maximum light, which occurs
about 3-4 days after the thermonuclear runaway process begins.}
\end{deluxetable}

\clearpage

\clearpage

\clearpage

\clearpage

\begin{figure}
\begin{center}
\includegraphics[width=4.5in]{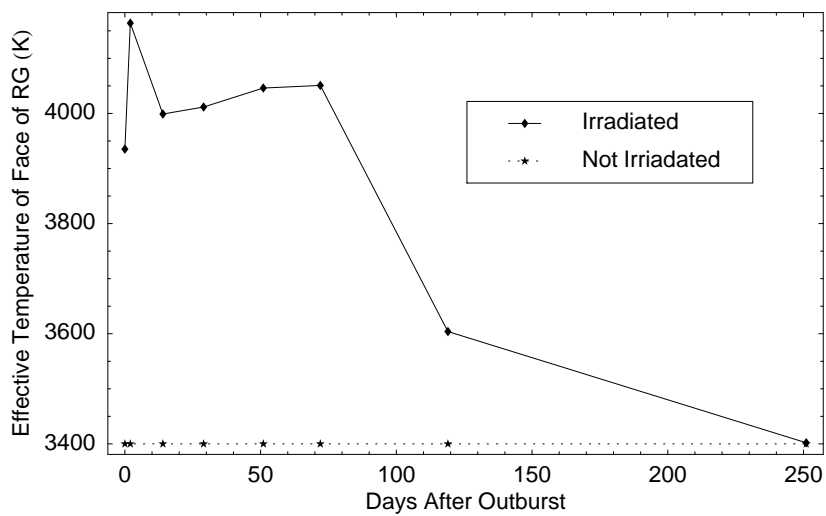}
\caption{Temperature of red giant star for the surface that is facing the Nova as a function of
the number of days post-outburst.  Also plotted is the temperature of the non-irradiated side of
the red giant. Modeled temperature rise is due to heating from irradiation by the nova and neglects any effect due to the passage of the forward shock. \label{RG-face-temp}}
\end{center}
\end{figure}


\clearpage
\begin{figure}
\begin{center}
\includegraphics[width=2.2in]{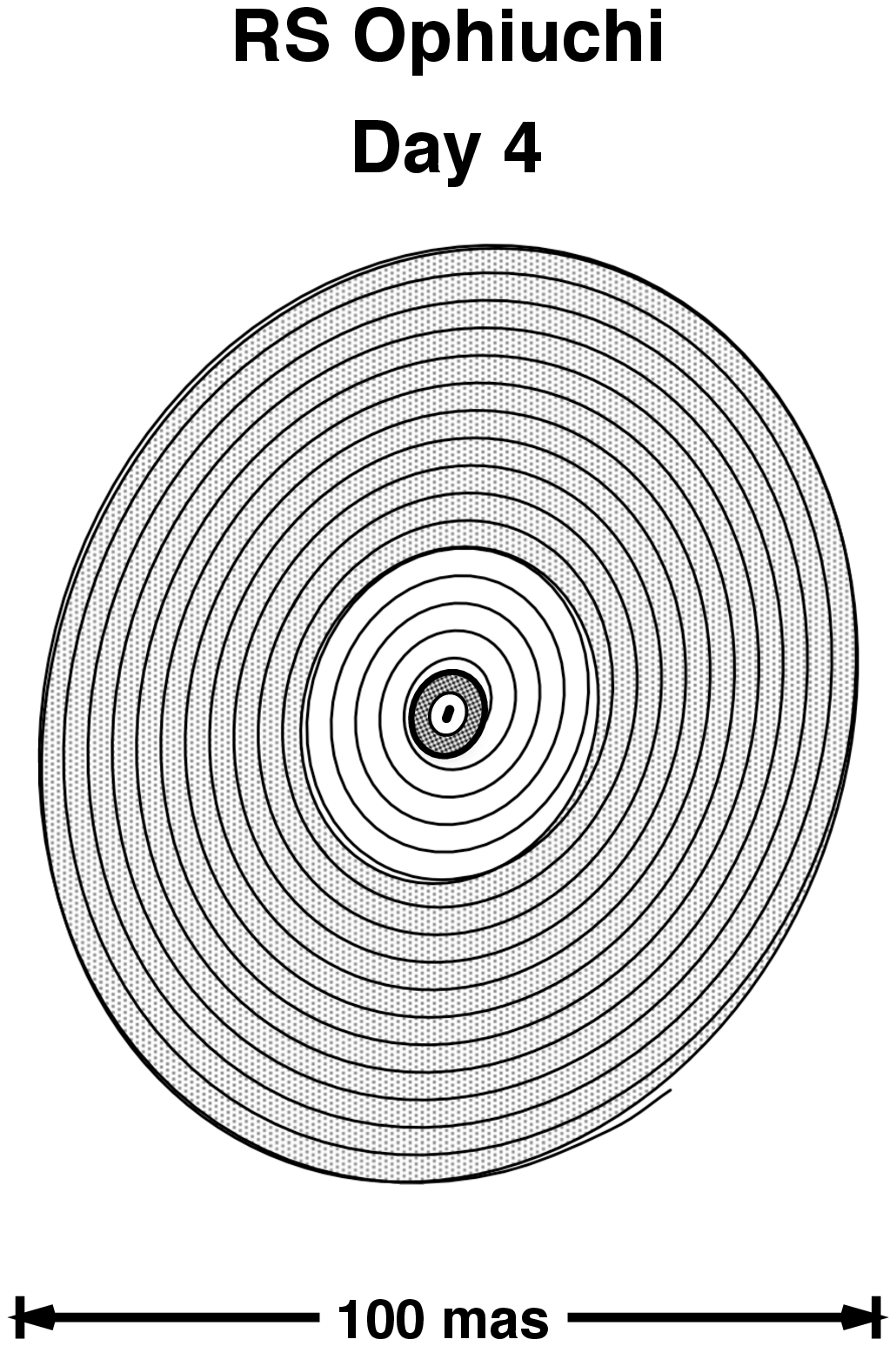}
\includegraphics[width=2.2in]{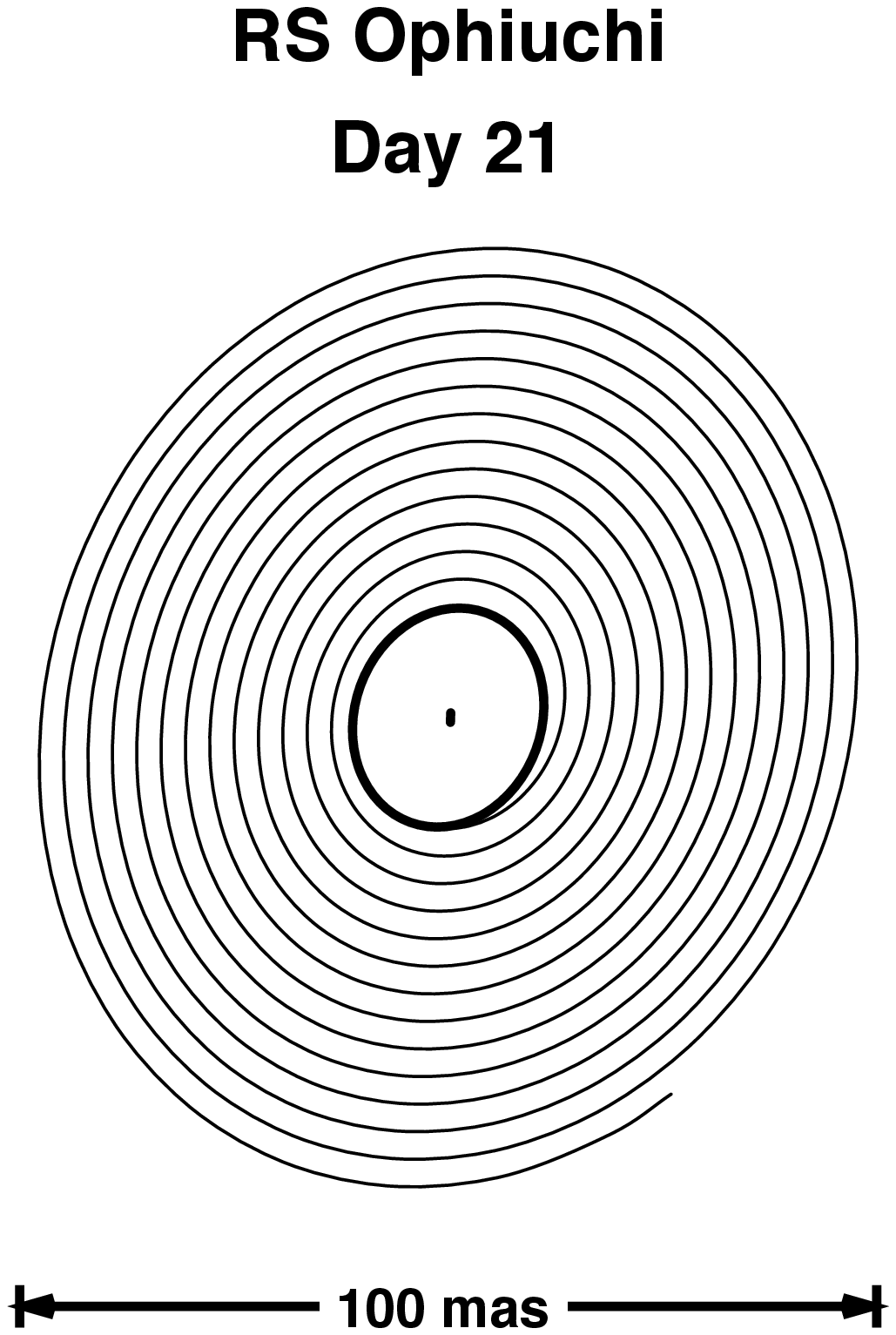}
\includegraphics[width=2.2in]{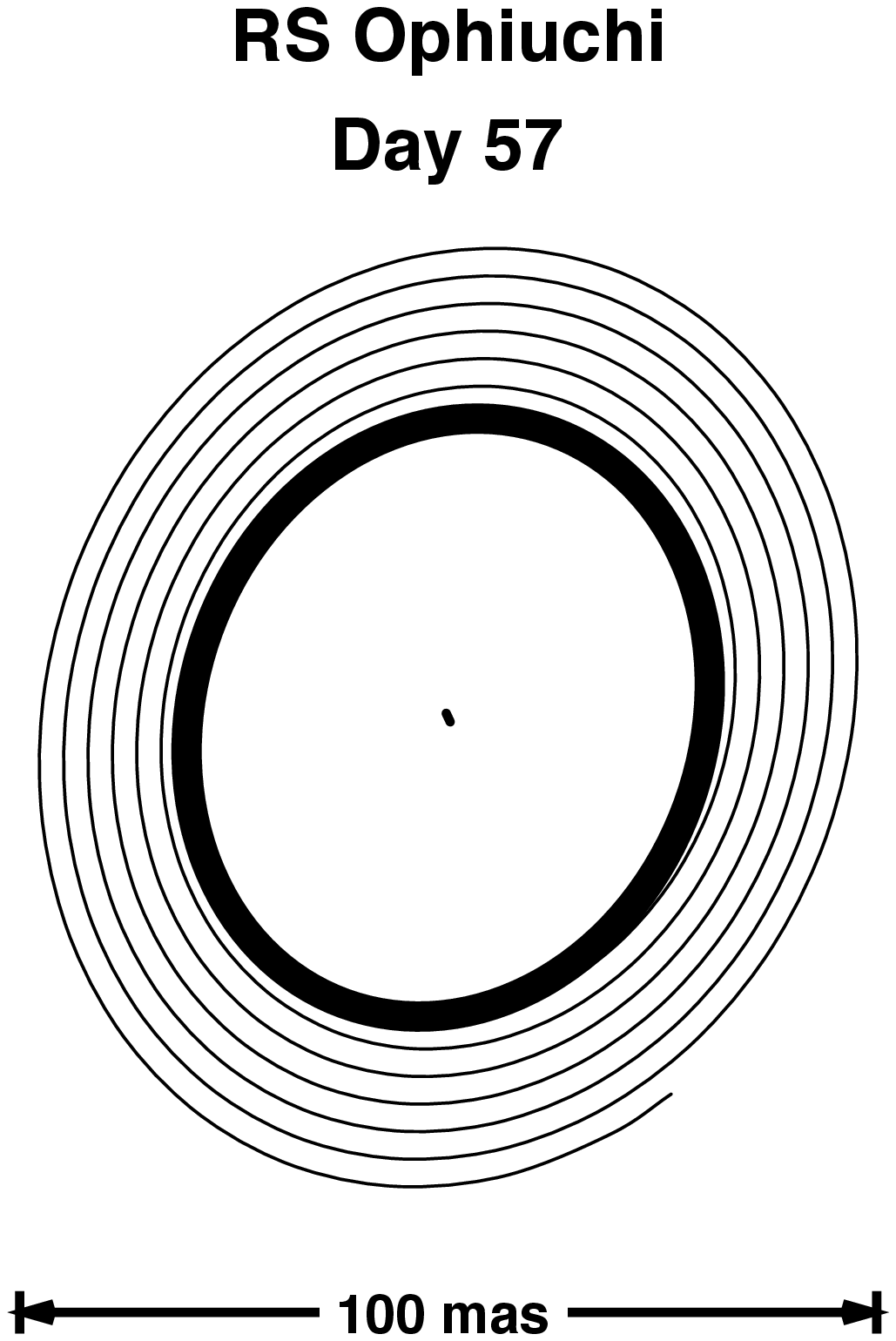}
\includegraphics[width=2.5in]{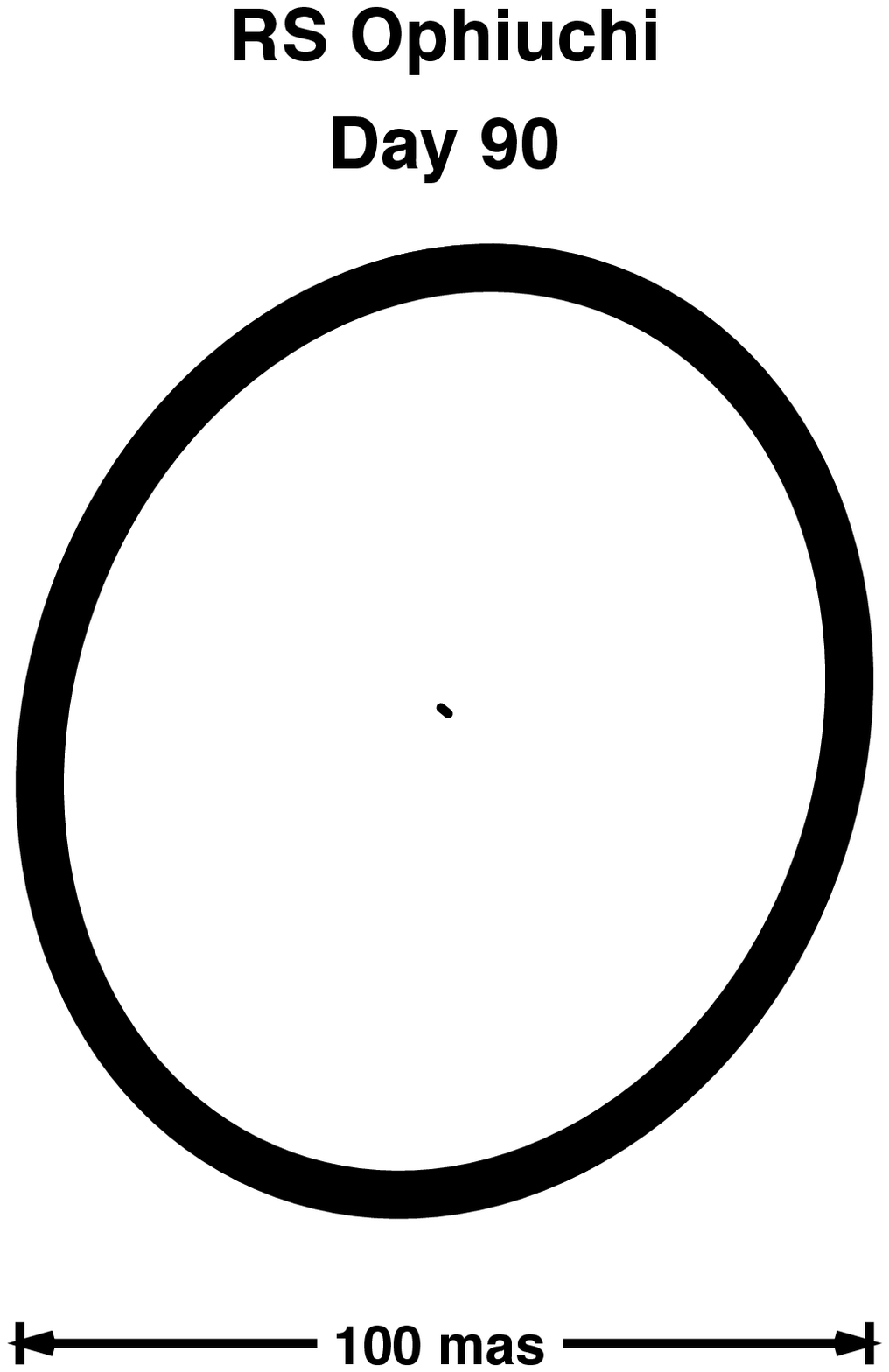}
\caption{Spiral shock wave model of RS Oph.     
(a) Top left panel displays the system geometry at 4 days post-outburst.  A gray ring is drawn in the
center of the figure to indicate the size of the shocked region at this epoch.  The outer part of the spiral
is overlayed with light gray to indicate that it is not known if the material stays in a coherent spiral
past the first several turns.  The diameter of the shocked region is about 5 mas assuming the
blast wave travels at a velocity of 1730 km/s in the plane of the orbit. 
(b) Top right panel.  The blast wave is now about 13 mas in diameter on day 21.
(c) Bottom left panel.  The blast wave is about 36 mas in diameter on day 57.
(d)  Bottom right panel.  By day 90 the blast wave has traversed the entire spiral pattern. \label{CSMatterEvol}}
\end{center}
\end{figure}


\begin{figure}
\begin{center}
\includegraphics[width=4.5in]{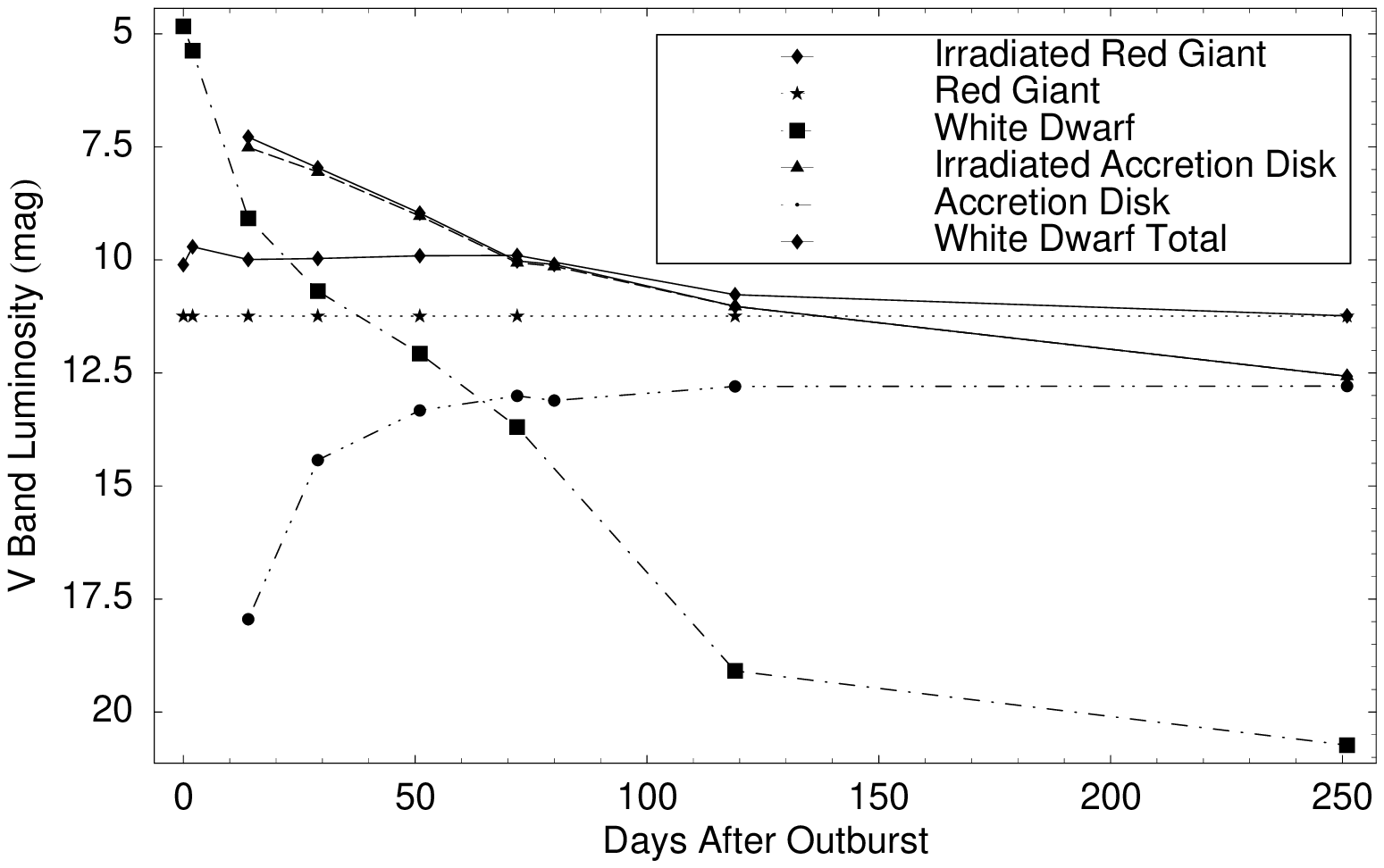}
\includegraphics[width=4.5in]{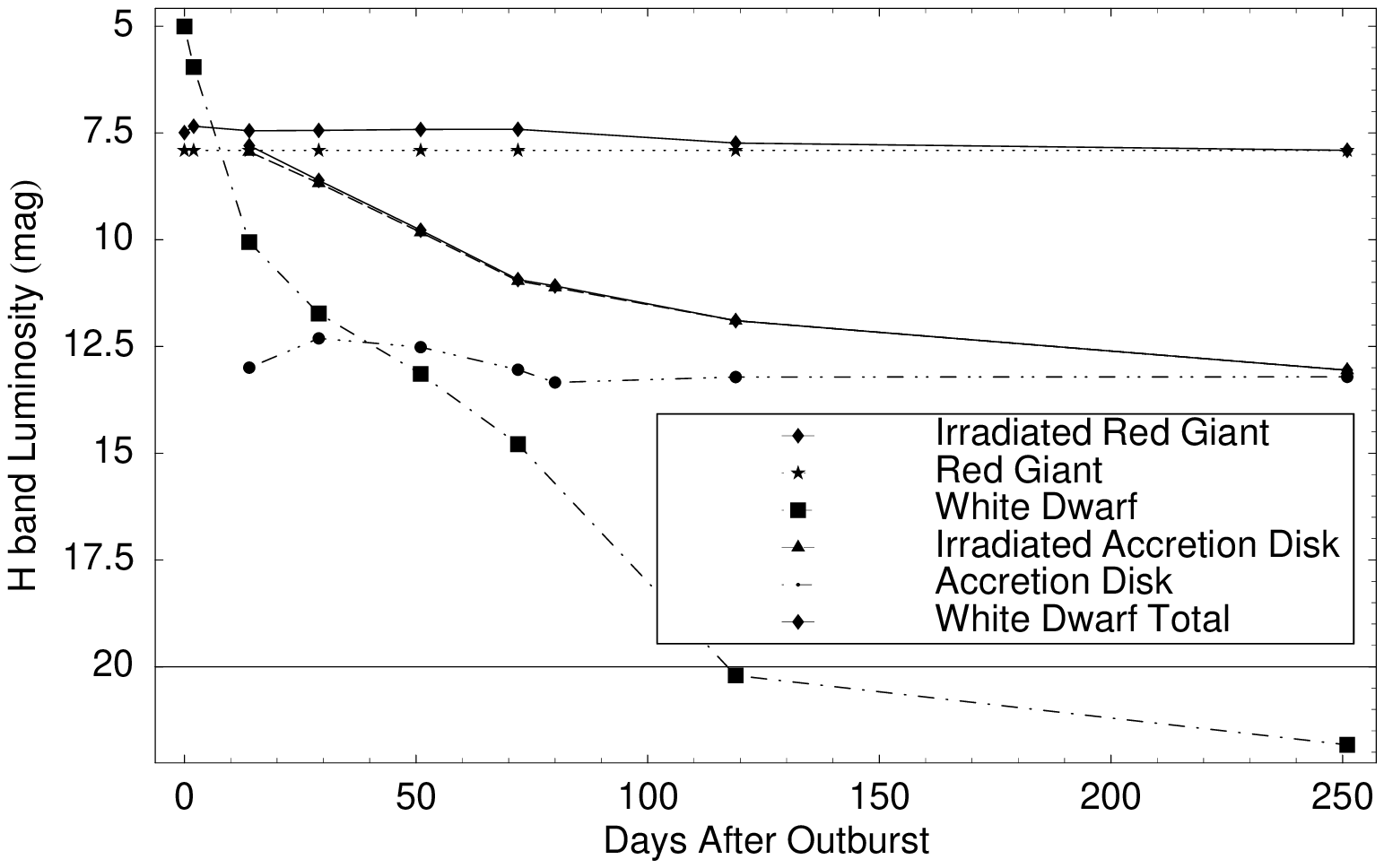}
\includegraphics[width=4.5in]{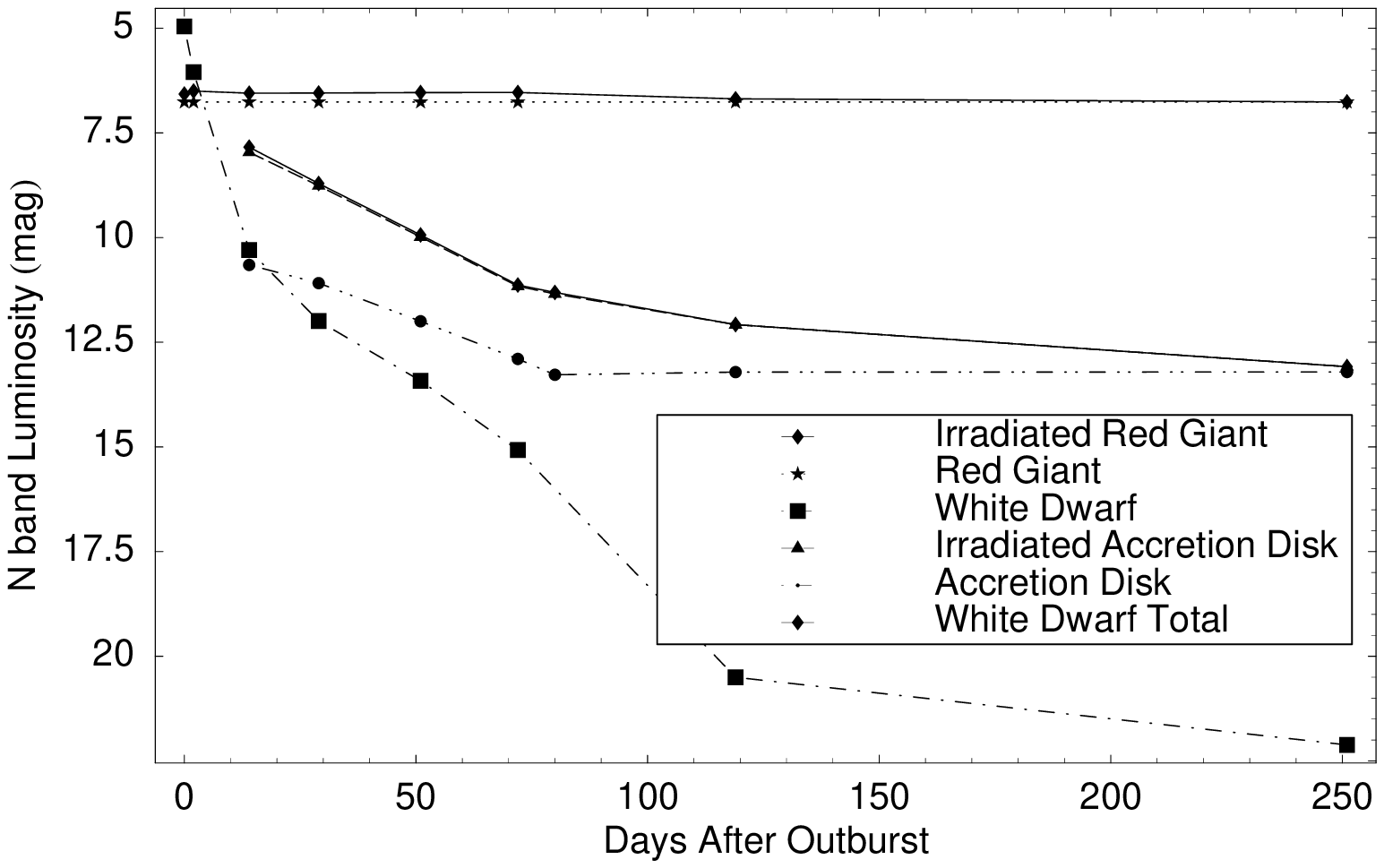}
\caption{(a) V Band luminosity as a function of time for Nova and red giant components of binary
system.(b) H band luminosity evolution.  (c) N band luminosity evolution.  \label{LumEvol}}
\end{center}
\end{figure}


\begin{thebibliography}{}

\bibitem[Adams \& Joy(1933)]{ada33}Adams, W. S., Joy, A. H., 1933, \pasp, 45, 249a

\bibitem[Barbon, Mammano \& Rosino(1969)]{bar69}Barbon, R., Mammano, A., Rosino, L., 1969, Comm. Konkoly Obs., 65, 257

\bibitem[Barry et al.(2008a)]{bar08} Barry, R.K., Mukai, K., Sokoloski, J. L., Danchi, W. C., Hachisu, I., Evans, A., Gehrz, R., \& Mikolajewska, J., 2008, in RS Ophiuchi (2006) and the recurrent nova phenomenon, eds A. Evans, M. F. Bode, T. J. O'Brien, Astronomical Society of the Pacific Conference Series, in press

\bibitem[Barry et al.(2008b)]{bar08b} Barry, R. K., Skopal, A., \& Danchi, W. C., 2008, \apj, manuscript in prep.

\bibitem[Blair et al.,(1983)]{bla83}Blair, W. P., Stencel, R. E., Feibelman, W. A., \& Michalitsianos, A. G., 1983, \apjs, 53, 573B

\bibitem[Bode \& Kahn(1985)]{bod85} Bode, M. F., \& Kahn, F. D., 1985, \mnras, 217, 205

\bibitem[Bode et al.(2006)]{bod06}Bode, M. F., O'Brien, T. J., Osborne, J. P., Page, K. L., Senziani, F., Skinner, G. K., Starrfield, S., Ness, J. -U., Drake, J. J., Schwarz, G., Beardmore, A. P., Darnley, M. J., Eyres, S. P. S., Evans, A., Gehrels, N., Goad, M. R., Jean, P., Krautter, J., \& Novara, G., 2006, \apj, 652, 629

\bibitem[Bode et al.(2007)]{bod07}Bode, M. F., Harman, D. J., O'Brien, T. J., Bond, H. E., Starrfield, S., Darnley M. J., Evans, A., \& Eyres, S. P. S., 2007, \apj, 665, L63

\bibitem[Brandi et al.(2008)]{bra08}Brandi, E., Quiroga, C., Ferrer, O. E., Miko\l ajewska, J., \& Garcia, L. G., 2008, in RS Ophiuchi (2006) and the recurrent nova phenomenon, eds A. Evans, M. F. Bode, T. J. O'Brien, Astronomical Society of the Pacific Conference Series, in press

\bibitem[Cardelli, Clayton \& Mathis(1989)]{car89}Cardelli, J. A., Clayton, G. C., \& Mathis, J. S., 1989, \apj, 345, 245

\bibitem[Cassatella et al.(1985)]{cas85}Cassatella, A., Harris, A. Snijders, M. A. J., \& Hassall, B. J. M., 1985, Proc. ESA Workshop: Recent Results on Cataclysmic Variables, ESA SP-236

\bibitem[Chesneau et al.(2007)]{che07}Chesneau, O., Nardetto, N., Millour, F., Hummel, C., Domiciano de Souza, A., Bonneau, D., Vannier, M., Rantakyro, F., Spang, A., Malbet, F., Mourard, D., Bode, M. F., O\'Brien, T. J., Skinner, G., Petrov, R. G., Stee, P., Tatulli, E., \& Vakili, F., 2007, \aap, 464, 119

\bibitem[Colavita et al.(2006)]{col06}Colavita, M.M. Serabyn, G. Wizinowich, P.L. \& Akeson, R.L. 2006, in Advances in Stellar Interferometry, eds. J.D. Monnier, M. Scholler and W.C. Danchi, Proc. SPIE 6268, 626803

\bibitem[Cowie \& Songaila(1986)]{cow86}Cowie, L. L., \& Songaila, A., 1986, \araa, 24, 499

\bibitem[Creech-Eakman et al.(2003)]{cre03}Creech-Eakman, M.J., Moore, J.D., Palmer, D.L., \& Serabyn, E., 2003, \procspie, 4841, 330c

\bibitem[Danchi \& Barry(2007)]{dan07} Danchi, W.C., \& Barry, R. K., 2007, \apj, manuscript in 
preparation

\bibitem[Das et al.(2006)]{das06} Das, R., Banerjee, D.P.K., \& Ashok, N.M. 2006, \apj, 653, 141

\bibitem[Dobrzycka \& Kenyon(1994)]{dob94}Dobrzycka, D., \& Kenyon, S. J., 1994, \apj, 108, 2259

\bibitem[Downes \& Duerbeck(2000)]{dow00}Downes, R. A., \& Duerbeck, H. W., 2000, \apj, 120, 2007

\bibitem[Evans et al.(2007a)]{eva07a}Evans, A., Kerr, T., Yang, B., Matsuoka, Y., Tsuzuki, Y., Bode, M. F., Eyres, S. P. S., Geballe, T. R., Woodward, C. E., Gehrz, R. D., Lynch, D. K., Rudy, R. J., Russell, R. W., O'Brien, T. J., Starrfield, S. G., Davis, R. J., Ness, J.-U., Drake, J., Osborne, J. P., Page, K. L., Adamson, A., Schwarz, G., \& Krautter, J. 2007, \mnras, 374, 1

\bibitem[Evans et al.(2007b)]{eva07b} Evans, A., Woodward, C.E., Helton, A., Gehrz, R. D., Lynch, D. K., Rudy, R. J., Russell, R. W., Bode, M. F., Kerr, T., Yang, B., Matsuoka, Y., Tsuzuki, Y.,  Eyres, S. P. S., Geballe, T. R., O'Brien, T. J.,  Davis, R. J., Starrfield, S. G., Ness, J.-U., Drake, J., Osborne, J. P., Page, K. L., Schwarz, G., \& Krautter, J., 2007, \apj, 663L, 29E

\bibitem[Evans et al.(2007c)]{eva07c}Evans, A., Woodward, C., Helton, A., van Loon, J. Th., Barry, R. K., Bode, M. F., et al., 2007, \apj, 671, L157  

\bibitem[Eyres et al.(1998)]{eyr98}Eyres, S. P. S., Evans, A., Salama, A, Barr, P., Clavel, J., Jenkins, N., Leech, K., Kessler, M., Lim, T., Metcalfe, L, \& Schulz, B., 1998, \apss, 255, 361

\bibitem[Fekel et al.(2000)]{fek00}Fekel, F. C., Joyce, R. R., Hinkle, K. H., \& Skrutskie, M. F., 2000, \apj, 119, 1375

\bibitem[Garcia(1986)]{gar86}Garcia, M.R., 1986, \aj, 91,1400

\bibitem[Grevesse(1984)]{gre84}Grevesse, N., 1984, Phys. Scr., T8, 49

\bibitem[Fleming(1904)]{fle04}Fleming, W., 1904, Harvard College Observatory Circular, 76

\bibitem[Hachisu \& Kato(2000)]{hac00}Hachisu, I., \& Kato, M., 2000, \apj, 536, 93

\bibitem[Hachisu \& Kato(2001)]{hac01}Hachisu, I., \& Kato, M., 2001, \apj, 558, 323

\bibitem[Hachisu et al.(2006)]{hac06}Hachisu, I., Kato, M., Kiyota, S., Kubotera, K., Maehara, H., et al., 2006, \apjl, 651, L141

\bibitem[Hartmann (1998)]{Hartmann98} Hartmann, L., 1998, in "Accretion Processes in Star Formation", [Cambridge UK, Cambridge University Press]

\bibitem[Hjellming et al.(1986)]{hje86}Hjellming, R. M., van Gorkom, J. H., Taylor, A. R., Seaquist, E. R., Padin, S., Davis, R. J., \& Bode, M. F., 1986, \apj, 305, 71

\bibitem[Kenyon \& Fernandez-Castro(1987)]{ken87}Kenyon, S. J., \& Fernandez-Castro, T., 1987, \apj, 93, 938

\bibitem[Kenyon \& Gallagher(1983)]{ken83}Kenyon, S. J., \& Gallagher, J. S., 1983, \aj, 88, 666K

\bibitem[Koresko et al.(2006)]{kor06}Koresko, C., Colavita, M., Serabyn, E., Booth, A., \& Garcia, J., 2006, \procspie, 6268, 626816-1

\bibitem[Lane et al.(2007)]{lan07}Lane, B. F., Sokoloski, J., Barry, R. K., Traub, W. A., Retter, A., et al., 2007, \apj, 658, 520

\bibitem[Lynch et al.(2006)]{lyn06}Lynch, D. K., Woodward, C. E., Geballe, T. R., Russell, R. W., Rudy, R. J., Venturini, C. C., Schwarz, G. J., Gehrz, R. D., Smith, N., Lyke, J. E., Bus, S. J., Sitko, M. L., Harrison, T. E., Fisher, S., Eyres, S. P., Evans, A., Shore, S. N., Starrfield, S., Bode, M. F., Greenhouse, M. A., Hauschildt, P. H., Truran, J. W., Williams, R. E., Perry, R. B., Zamanov, R., \& O'Brien, T. J., 2006, \apj, 638, 987

\bibitem[Lynden-Bell \& Pringle(1974)]{Lynden74} Lynden-Bell, D. \& Pringle, 1974, MNRAS, 168, 603

\bibitem[Mastrodemos \& Morris(1999)]{mas99}Mastrodemos, N., \& Morris, M., 1999, \apj,
523, 357

\bibitem[Mauron \& Huggins(2006)]{mau06}Mauron, N., \& Huggins, P.J., 2006, \aap, 452, 257

\bibitem[Monnier et al.(2006)]{mon06}Monnier, J. D., Barry, R. K., Traub, W. A., Lane, B. F., Akeson, R. L., Ragland, S., Schuller, P. A., Berger, J. P., Millan-Gabet, R., Pedretti, E., Schloerb, F. P., Koresko, C., Carleton, N. P., Lacasse, M. G., Kern, P., Malbet, F., Perraut, K., \& Muterspaugh, M. W., 2006, \apjl, 647, 127

\bibitem[Morris et al.(2006)]{mor06}Morris, M., Sahai, R., Matthews, K., Cheng, J., Lu, J., 
Claussen, M., \& Sanchez-Contreras, C., 2006,  in Planetary Nebulae in our Galaxy and Beyond,
Proceedings of IAU Symposium No. 234, M. J. Barlow \& R. H. Mendez, eds., 469.

\bibitem[Morrison(1985)]{mor85}Morrison, W., 1985, \iaucirc, 4030

\bibitem[M\"{u}rset \& Schmid(1999)]{mur99} M\"{u}rset, U., \& Schmid, H.M., 1999, \aaps, 137, 473

\bibitem[Narumi(2006)]{nar06}Narumi, H., Hirosawa, K., Kanai, K., Renz, W., Pereira, A.,Nakano, S., Nakamura, Y., \& Pojmanski, G. 2006, \iaucirc, 8671, 2

\bibitem[O'Brien et al.(2006)]{obr06}O'Brien, T. J., Bode, M. F., Porcas, R. W., Muxlow, W. B., Eyres, S. P. S., Beswick, R. J., Garrington, S. T., Davis, R. J., \& Evans, A., 2006, \nat, 442, 279

\bibitem[Oppenheimer \& Mattei(1993)]{opp93}Oppenheimer, B.D., \& Mattei, J.A., 1993, JAVSO, 22, 1050

\bibitem[Payne-Gaposchkin(1957)]{pay57}Payne-Gaposchkin, C., 1957, The Galactic Novae, North-Holland, Amsterdam


\bibitem[Pottasch(1967)]{pot67}Pottasch, S. R., 1967, \bain, 19, 227 %

\bibitem[Pringle (1981)]{Pringle81} Pringle, J. E. 1981, ARA\&A, 19, 137

\bibitem[Rupin et al.(2008)]{rup08}Rupin, M. P., Mioduszweski, A. J., Sokoloski, J. L., Kaiser, C. R., \& Brocksopp, C., 2006, Proc. AAS 

\bibitem[Sacuto et al.(2007)]{sac07}Sacuto, s., Chesneau, O., Vannier, M., \& Cruzalebes, P., 2007, \aaps, 465, 469

\bibitem[Schaeffer(2004)]{sch04}Schaeffer B., 2004, IAUC 8396

\bibitem[Schild et al.(2001)]{sch01}Schild, H., Eyres, S. P. S., Salama, A., \& Evans, A., 2001, \aaps, 378, 146

\bibitem[Serabyn et al.(2000)]{ser00}Serabyn, E., Colavita, M.M. \& Beichman, C.A. 2000, in Thermal Emission
Spectroscopy and Analysis of Dust, Disks, and Regoliths, ASP Conf. Ser. Vol. 196, eds. M.L. Sitko, A.L. Sprague \& D.K. Lynch, p. 357.

\bibitem[Serabyn et al.(2004)]{ser04}Serabyn, E. et al. 2004, in SPIE Vol. 5491, New Frontiers in Stellar
Interferometry, ed. W.A. Traub, 136.

\bibitem[Serabyn et al.(2005)]{ser05}Serabyn, E. et al. 2005, in SPIE Vol. 5905, Techniques and Instrumentation for Detection of Exoplanets II, ed. D.R. Coulter, 5905OT-1.

\bibitem[Serabyn et al.(2006)]{ser06}Serabyn, E. et al. 2006, Proc. SPIE 6268, Advances in Stellar Interferometry, eds. J.D. Monnier, M. Scholler and W.C. Danchi, 626815.

\bibitem[Serabyn et al.(2008)]{ser08}Serabyn, E., et al., 2008, \apj, manuscript in preparation

\bibitem[Snijders   (1985)]{sni85}Snijders, M. A. J., 1985, \apss, 130, 244

\bibitem[Sokoloski et al.(2006)]{sok06}Sokoloski, J. L., Luna, G. J. M., Mukai, K., \& Kenyon, S. J., 2006, \nat, 442, 276

\bibitem[Starrfield, Sparks \& Truran(1985)]{sta85}Starrfield, S., Sparks, W. M., \& Truran, J. W., 1985, \apj, 291, 136

\bibitem[Wallerstein(1958)]{wal58}Walerstein, G., 1958, \pasp, 70, 537w

\bibitem[Warner (1995)]{war95} Warner, B., 1995, Cataclysmic Variable Stars, [Cambridge, UK: Cambridge University Press], p. 27.

\end{thebibliography}
\end{document}